\documentclass[useAMS,usedcolumn,usenatbib,usegraphicx]{mn2e}
\usepackage{times}

\usepackage{aas_macros}
\usepackage{textcmds}
\usepackage{multirow}
\usepackage{xcolor}
\usepackage{hyperref}
\usepackage{ctable}
\usepackage{amsmath,amssymb}
\usepackage{diagbox, lscape, chemfig, subcaption, textcomp}
\newcommand\T{\rule{0pt}{2.6ex}}       % Top strut
\newcommand\B{\rule[-1.2ex]{0pt}{0pt}} % Bottom strut
\DeclareSymbolFont{matha}{OML}{txmi}{m}{it}% txfonts
\DeclareMathSymbol{\varv}{\mathord}{matha}{118}
\title[Ingredients for Solar-like Systems]{Ingredients for Solar-like Systems: protostar IRAS~16293-2422~B versus comet 67P/Churyumov--Gerasimenko}

\author[Maria N. Drozdovskaya et al.]{Maria~N.~Drozdovskaya$^{1}$\thanks{E-mail: maria.drozdovskaya@csh.unibe.ch}, Ewine~F.~van~Dishoeck$^{2,3}$, Martin Rubin$^{4}$,\newauthor
Jes~K.~J{\o}rgensen$^{5}$, Kathrin Altwegg$^{4}$\\
$^{1}$~Center for Space and Habitability, Universit\"{a}t Bern, Sidlerstrasse 5, 3012 Bern, Switzerland\\
$^{2}$~Leiden Observatory, Leiden University, P.O. Box 9513, 2300 RA, Leiden, The Netherlands\\
$^{3}$~Max-Planck-Institut f\"{u}r Extraterrestrische Physik, Giessenbachstrasse 1, 85748 Garching, Germany\\
$^{4}$~Physikalisches Institut, Universit\"{a}t Bern, Sidlerstrasse 5, 3012 Bern, Switzerland\\
$^{5}$~Niels Bohr Institute \& Centre for Star and Planet Formation, University of Copenhagen, {\O}ster Voldgade 5--7, 1350 Copenhagen K., Denmark
}

\begin{document}

\date{Accepted xxx.  Received xxx; in original form xxx}

\pagerange{\pageref{firstpage}--\pageref{lastpage}} \pubyear{2019}

\maketitle
\label{firstpage}

\begin{abstract}
Our modern day Solar System has $4.6\times10^{9}$~yrs of evolution behind it with just a few relics of its birth conditions remaining. Comets are thought to be some of the most pristine tracers of the initial ingredients that were combined to produce the Earth and the other planets. Other low-mass protostars may be analogous to our proto-Sun and hence, could be used to study the building blocks necessary to form Solar-like systems. This study tests this idea on the basis of new high sensitivity, high spatial resolution ALMA data on the protoplanetary disc-scales ($\sim70$~au) of IRAS~16293-2422 and the bulk composition of comet 67P/Churyumov-Gerasimenko, as determined for the first time with the unique in situ monitoring carried out by \textit{Rosetta}. The comparative analysis of the observations from the Protostellar Interferometric Line Survey (PILS) and the measurements made with Rosetta Orbiter Spectrometer for Ion and Neutral Analysis (ROSINA) shows that the relative abundances of CHO-, N-, and S-bearing molecules correlate, with some scatter, between protostellar and cometary data. A tentative correlation is seen for the first time for P- and Cl-bearing compounds. The results imply that the volatile composition of cometesimals and planetesimals is partially inherited from the pre- and protostellar phases of evolution.
\end{abstract}


%%%%%%%%%%%%%%%%%%%%%%%%%%%%%%%%%%%%%%%%%%%%%%%%%%%%%%%%%%%%%%%%%%%%%%%%%%%%%%%
\clearpage
\newpage
\section{Introduction}
\label{introduction}

Frozen volatile molecules are found in our Solar System in cold distant regions from the Sun or within bodies sufficiently large to shelter the ices from thermal desorption. Consequently, this includes large ($<10^{3}$ km) icy moons such as Europa or Enceladus, and small ($\sim$km-sized) distant comets. As our mature Solar System is devoid of gas on disc-scales, ices still present today must have been formed during earlier evolutionary phases of our system when gases were still available for adsorption. This implies that cometary ices are made from the gases and ices found in the protoplanetary disc and the prestellar core \citep{GreenbergLi1999, EhrenfreundCharnley2000}. Prestellar ices may be entirely inherited by comets (i.e., pristine), or may be partially or entirely modified en route to and inside the disc and into the comets (i.e., partial or full reset). Recent measurements of a very high ratio of $17$ for D$_{2}$O/HDO relative to HDO/H$_{2}$O in comparison to the statistically expected value of $0.25$ \citep{Altwegg2017a} on comet 67P/Churyumov--Gerasimenko, hereafter 67P/C--G, corroborate the pristinity of cometary water ice. The abundance of highly volatiles species, such as CO, N$_{2}$ and noble gases \citep{Rubin2018} on 67P/C--G rejects the possibility of full reset, as the forming disc is unlikely to ever be sufficiently cold to re-adsorb these molecules. Comets and other icy planetesimals have been postulated to bring water and the ingredients for life to our planet due to their significant late-time dynamics. Hence, understanding their composition and origins may shed light on the emergence of life on Earth, its ubiquity on other planets and in extrasolar systems (as reviewed in \citealt{MummaCharnley2011, AHearn2011b, Bockelee-Morvan2015a}).

The formation of a protostar and its protoplanetary disc is governed by gravitational collapse \citep{Shu1987}. This process is coupled with grain-growth mechanisms transforming $0.1$~$\mu$m-sized dust grains found in prestellar cores to mm-sized dust particles seen in discs. Disc-scale gas and dust hydrodynamic processes subsequently assemble m-sized planetesimals. Cometary bodies may be a normal by-product of planet formation across the disc in the form of remnant building blocks or even primordial rubble piles composed of their own building blocks \citep{AHearn2011b, Davidsson2016}. Once the icy dust/rocks are assembled into a km-sized body, it is unlikely to be significantly thermally processed. Most recent calculations suggest that even a perihelion passage of a comet at $1.2$~au from the Sun will only heat the outermost few tens of cm (at least for the morphology of 67P/C--G; \citealt{Schloerb2015, Capria2017}). Meanwhile, non-catastrophic collisions are also unlikely to generate sufficient energy for significant heating (e.g., \citealt{Jutzi2017, JutziBenz2017, Schwartz2018}). Hence, bulk cometary ices very likely closely resemble disc and prestellar ices and gases \citep{Pontoppidan2014}. Consequently, cometary bulk composition may yield information about the ingredients for Solar-like systems.

For a long time, the Oort cloud comet C/1995~O1 (Hale--Bopp) was the best studied cometary body thanks to its high brightness (total visual magnitude of $\sim 10.5$ at a heliocentric distance of $7$~au). \citet{Bockelee-Morvan2000} used ground-based sub-mm ($80-370$~GHz) facilities to study the chemical composition of Hale--Bopp's coma between February and April 1997 for heliocentric distances in the $0.91-1.2$~au range close to its perihelion on April 1, 1997. These data were used to infer a strong correlation between Hale--Bopp's abundances of CHO- and N-bearing molecules and those derived from interstellar medium (ISM) observations on envelope- or cloud-scales (thousands of au). In contrast, the S-bearing species showed a large scatter. At the time, it was unclear whether such trends would persist for other comets. Now, the \textit{Rosetta} mission has yielded unprecedented detail and wealth of  information on another Jupiter-family comet, 67P/C--G (as reviewed in Altwegg et al. in press). The mission accompanied and continuously monitored the comet with its suite of instruments for more than $2$ years pre- and post- its August 12, 2015 perihelion for heliocentric distances starting at $4.4$~au down to $1.24$~au and back out to $3.8$~au (distances of aphelion and perihelion are $5.683$ and $1.243$~au, respectively).

The target of the \textit{Rosetta} mission, comet 67P/C--G, is composed of two lobes that are $4.1\times3.5\times1.6$ and $2.5\times2.1\times1.6$~km in size \citep{Jorda2016}. Thanks to the continuous monitoring of the target by the orbiter, it was realized that there is much variability in the outgassing of ices hidden underneath the surface, which has to do with seasonal and diurnal variations as probed with Rosetta Orbiter Spectrometer for Ion and Neutral Analysis (ROSINA), Visible and Infrared Thermal Imaging Spectrometer (VIRTIS), Microwave Instrument for the Rosetta Orbiter (MIRO) and Optical, Spectroscopic, and Infrared Remote Imaging System (OSIRIS) instruments aboard \textit{Rosetta} (e.g., \citealt{DeSanctis2015, Lee_S2015, Bockelee-Morvan2015b, Biver2015b, Luspay-Kuti2015, Hassig2015, Filacchione2016a, Filacchione2016b, Fornasier2016, Hansen2016, Bockelee-Morvan2016, Barucci2016, Migliorini2016, Gasc2017, Marshall2017, Filacchione2019}). The bi-lobate geometry of the nucleus and the associated self-shielding, its changing rotational period, backfall of granular material, short-lived outbursts, active sinkhole pits and orbital trajectory change the irradiance of its surface during a single apparition \citep{Keller2015, Vincent2015, Vincent2016, Feldman2016, Keller2017, Kramer2018}, but also in the long term upon repeated approaches to the Sun. Nevertheless, it is possible to extract bulk abundances of the interior ices upon careful data analysis (e.g., \citealt{Calmonte2016}) and to peek at them on special occasions such as cliff collapses \citep{Pajola2017}. The nucleus of 67P/C--G is thought to be homogeneous based on Comet Nucleus Sounding Experiment by Radiowave Transmission (CONSERT) and Radio Science Investigation (RSI) experiment measurements \citep{Kofman2015, Patzold2016}. It seems likely that its shape stems from a merger of two distinct objects (e.g., \citealt{Massironi2015, JutziAsphaug2015, Matonti2019}).

ISM astrochemical studies have also been profiting from new facilities capable of high spatial resolution observations, allowing Solar System-scales ($<100$~au) to be probed for the first time in young star-forming regions (e.g., \citealt{ALMA2015, Ansdell2016}). The closest Solar-like system that is still in its infant embedded phase of formation, the low-mass binary IRAS~16293-2422, has been the target of many observational campaigns over the last few decades with single-dish and interferometric facilities (see \citealt{Jorgensen2016} for a review). The system still finds itself in the earliest, gas-rich, embedded phase of star formation corresponding to the suspected time of cometesimal formation. Now, with ALMA, accurate abundances of several tens of molecules can be determined in the disc-like structures found in this puerile system. This allows the volatiles being incorporated into comets and planetesimals in IRAS~16293-2422 to be directly compared to those found in our Solar System on analogous spatial scales. Early hints for the interstellar -- cometary relation in the context of this protostellar source and comet Hale--Bopp have been investigated by \citet{Schoier2002}, but were unable to access the most inner disc-forming regions based on data from facilities less powerful in comparison to ALMA.

IRAS~16293-2422 is the closest protostellar low-mass system that has been well-characterized physically and chemically \citep{Jorgensen2016}. It is composed of two deeply embedded protostars, A and B, at a short distance of $141$~pc \citep{Dzib2018} with a projected separation of $5\farcs{3}$ ($747$~au). The masses and luminosities have been estimated to be on the order of $18$~$L_{\sun}$, $1.0$~$M_{\sun}$ for source A and $3$~$L_{\sun}$, $0.1$~$M_{\sun}$ for source B, based on previous observations and theoretical models \citep{Jacobsen2018a}. The total amount of mass encompassed by the circumbinary envelope of $\sim50\arcsec$ in size is $\sim4$~$M_{\sun}$ \citep{Jacobsen2018a}. High spatial resolution observations with ALMA have resolved the scales of the two individual discs, i.e., on scales of a few tens of au. The data suggest that the disc around source A is nearly edge-on, while that around source B is face-on (e.g., \citealt{Pineda2012, Zapata2013}). This has also been independently supported via dust continuum polarization studies (e.g., \citealt{Liu2018, Sadavoy2018}). The outer dust disc radius of B is suggested to be about $30-56$~au \citep{Rodriguez2005, Zapata2013, Hernandez-Gomez2019b}. The velocity gradient across the `disc'-domain of source B \citep{Zapata2013} is much shallower than that across A \citep{Girart2014}. It has so-far not been possible to determine the relative ages of the A and B sources using signatures of infall and chemical differentiation (e.g., \citealt{Chandler2005, Zapata2013, Calcutt2018a, Rivilla2019, vanderWiel2019}). It is unlikely for there to be a drastic age gap, as the two protostars are part of a binary system and are both still undergoing gravitational collapse.

To explore the hypothesis of close ties between cometary and protostellar chemical inventories, it is necessary to assume that all low-mass systems evolve analogously. In this paper, comet 67P/C--G will be considered as a representative probe of the bulk cometary ices; and IRAS~16293-2422~B will assume the role of a Solar-like embedded system. The goal of this paper is to compare the chemical inventories of these two targets and thereby test the chemical links that may or may not exist between cometary and interstellar volatiles. This work showcases the synergy of the powerful capabilities of \textit{Rosetta} and ALMA. Section~\ref{methods} describes the data that are used in this paper to obtain the results presented in Section~\ref{results}. The implications of the findings are presented in Section~\ref{discussion} and the conclusions are summarized in Section~\ref{conclusions}.

%%%%%%%%%%%%%%%%%%%%%%%%%%%%%%%%%%%%%%%%%%%%%%%%%%%%%%%%%%%%%%%%%%%%%%%%%%%%%%%
\section{Methods}
\label{methods}

\subsection{67P/C--G}
\label{67P}

The data on comet 67P/C--G analysed in this work stem from the ROSINA instrument suite aboard the orbiter, which measures the gases stemming from the comet at the distance of the orbiter from the comet surface. This reduces the uncertainties stemming from photodissociation rates, which are required to correct for photodissociation of molecules in the coma when observing with ground-based facilities \citep{Bockelee-Morvan2000}. The ROSINA Double Focusing Mass Spectrometer (DFMS) has a high mass resolution ($m/\Delta m = 3000$ on mass/charge of $28$~u/e at the $1$ per cent peak height) and the ROSINA Reflection-type Time-Of-Flight (RTOF) mass spectrometer has a wide mass range ($1-1000$~u/e), allowing unambiguous identification of small and large molecules \citep{Balsiger2007}.

Here, the averages of the measurements obtained between the 22nd of May and the 2nd of June, 2015 are used as bulk abundances. During this period, the orbiter was at distances in the $100-200$~km range from the comet surface. This specific May 2015 time frame is ideal for measuring the bulk volatile content (\citealt{Calmonte2016}; Altwegg et al. in press). It starts at the final pre-2015-perihelion equinox of the comet and ends prior to the coma becoming significantly polluted with dust as a result of higher activity closer to the Sun. During this period, the Southern hemisphere began experiencing the summer season (which is short, yet intense, in comparison to the summer experienced by the Northern hemisphere). This hemisphere is thought to be less covered by the resettled dust from earlier perihelia \citep{Keller2017}. Finally, during this time, 67P/C--G was within $\sim2$~au from the Sun, thus subjecting it to surface temperatures well above those required for thermal desorption of water. This implies that almost all the volatiles were sublimating at this time, unlike at larger distances when the volatility of molecules affects their observed desorption patterns. Even closer to the Sun, coma abundances become affected by outbursts, which appear to be powered by CO$_{2}$ and in turn, are less representative of the bulk interior.

\subsection{IRAS~16293-2422}
\label{IRAS16293}

In this work, the majority of observational data on IRAS~16293-2422 stem from the large unbiased Protostellar Interferometric Line Survey (PILS\footnote[2]{\url{http://youngstars.nbi.dk/PILS/}}; project-id: 2013.1.00278.S, PI: Jes K. J{\o}rgensen) carried out with ALMA in the $329-363$~GHz frequency range (Band $7$) during Cycle 2 observations, supplemented with ALMA observations in Bands $3$ ($\sim100$~GHz) and $6$ ($\sim230$~GHz) carried out during Cycle 1 (project-id: 2012.1.00712.S, PI: Jes K. J{\o}rgensen; \citealt{Jorgensen2016}). This dataset represents the most complete spectral characterization of the source on (almost) identical spatial scales (the beam size does vary slightly with frequency across the large range observed). PILS was carried out at a spectral resolution of $0.2$~km~s$^{-1}$ ($0.244$~MHz) and restored with a uniform circular beam of $0\farcs{5}$. The data from the main array of $12$~m dishes are combined with data from the Atacama Compact Array (ACA) of $7$~m dishes, hence, resulting in the largest recoverable size of $13\arcsec$. This implies that the PILS dataset can be used to study the emission on the scale of the individual envelopes of the two protostars ($\sim1-3\arcsec$) and on the scale of their discs ($<1\arcsec$). The main position analysed in this work is a one-beam offset position from source B in the SW direction, which optimizes high densities (hence, boosting column densities of the least abundant molecules), while avoiding the self-absorption and dust absorption at the highest available densities found on-source (e.g., \citealt{Coutens2016, Lykke2017, Ligterink2017, Calcutt2018a, Drozdovskaya2018, Manigand2019}). Relative abundances at the half-beam offset position from source B in the same direction are similar to those at the full-beam offset position, but with a factor of $2$ higher column density \citep{Jorgensen2018}. The offset positions in terms of the continuum and molecular emission distributions have been shown in previous publications (e.g., fig.~$7$ of \citealt{Jorgensen2016}; fig.~$2$ of \citealt{Jorgensen2018}; or fig.~$1$ of \citealt{Drozdovskaya2018}). The narrower line widths associated with the dynamical structure near source B reduce line blending. Taking the known physical structure of source B and its disc into account, at $0\farcs{5}$ ($\sim70$~au) from the source, the observations probe the material entering the protoplanetary disc, especially, since infalling velocity signatures are seen \citep{Pineda2012}.

\clearpage
\ctable[
 caption = {Quantities of volatiles towards IRAS~16293-2422~B as observed with ALMA on protoplanetary disc-scales\tmark.},
 label = {tbl:abunvalues_B}
 ]{@{\extracolsep{\fill}}lllll}{
 \tnote{\tiny{The assumed source size is $0\farcs{5}$; and $\eta_{\text{BF}}=\text{source size}^{2}/(\text{source size}^{2}+\text{beam size}^{2})$, when the source size is smaller than the beam size; and $\eta_{\text{BF}}=\text{source size}^{2}/\text{beam size}^{2}$, when the beam size is smaller than the source size.}}
\tnote[b]{\tiny{A correction factor for the coupling of line emission with the emission from dense dust at $T_{\text{bg}}$ needs to be accounted for when deriving the column density at the one-beam offset position from source B. \citep{Persson2018} used a consistently derived value of $1.1658$. \citet{Ligterink2017, Fayolle2017} and Manigand et al. (subm.) also consistently corrected for $T_{\text{bg}}=21$~K. \citet{Jorgensen2016, Jorgensen2018, Calcutt2018a, Calcutt2018b} applied correction factors of $1.14$ and $1.05$ for $T_{\text{ex}}=125$~K and $300$~K, respectively (corresponding to $T_{\text{bg}}=21$~K). \citet{Coutens2016, Lykke2017, Drozdovskaya2018, Taquet2018} did not account for this factor in the published column densities, and hence, these values have been corrected in this work through division by $1.14$ or $1.05$ depending on the $T_{\text{ex}}$ of the molecule in question. This has either been explicitly stated by the authors or has been conveyed via private communication. For the column densities of PO, PN, and glycine that are newly derived in this work, these correction factors have been applied. For the column densities of CO, HCN and HNC that are newly derived in this work, these correction factors are not applied due to highly uncertain spatial distributions of these molecules. The accuracy of the derived column densities is $10-20$ per cent. Variations in $T_{\text{ex}}$ of $\sim20$ per cent ($25-60$~K for $T_{\text{ex}}=125-300$~K) change the derived column densities by $<10$ per cent \citep{Jorgensen2018}.}}
 \tnote[c]{\tiny{For a $\text{beam size}=0\farcs{2}$, $N(\text{H}_{2}^{18}\text{O})=9.5\times10^{17}$~cm$^{-2}$. Assuming $^{16}\text{O}/^{18}\text{O}=557$ and using $\eta_{\text{BF}}\approx6$, gives $N(\text{H}_{2}\text{O})=N(\text{H}_{2}^{18}\text{O})\times\eta_{\text{BF}}\times(^{16}\text{O}/^{18}\text{O})$ as the value used in this work for a source size of $0\farcs{5}$ under the assumption of homologous emission.}}
 \tnote[d]{\tiny{The published column densities have been reduced by a factor of $2.136$ to account for the shift from the half-beam offset position to the full-beam offset position from source B.}}
 \tnote[e]{\tiny{Several estimates exist in the literature, but all are only for the circumbinary envelope. \citet{Hily-Blant2010}: $T_{\text{ex}}=8-10$~K, $\text{beam size}=12-18\arcsec$, $N(\text{NH}_{3})=20-3.5\times10^{15}$~cm$^{-2}$. \citet{vanDishoeck1995}: $T_{\text{ex}}=25\pm5$~K, $\text{beam size}=20\arcsec$, $N(\text{NH}_{2}\text{D})=1.9\times10^{14}$~cm$^{-2}$; assuming $\text{D}/\text{H}=0.05-0.005$ gives $N(\text{NH}_{3})=N(\text{NH}_{2}\text{D})\times(\text{H}/\text{D})=3.8-38\times10^{15}$~cm$^{-2}$. \citet{Mundy1990}: $T_{\text{ex}}=15-20$~K, $\text{beam size}=20\arcsec$, $N(\text{NH}_{3})=2\times10^{15}$~cm$^{-2}$. \citet{Mizuno1990}: $T_{\text{ex}}=15$~K, $\text{beam size}=40\arcsec$, $N(\text{NH}_{3})=8\times10^{14}$~cm$^{-2}$; assuming $\text{source size}=20\arcsec$ yields $\eta_{\text{BF}}=0.2$ and gives $N(\text{NH}_{3})/\eta_{\text{BF}}=4\times10^{15}$~cm$^{-2}$. Taking all these estimates for a $\text{beam size}=20\arcsec$ yields a range: $N(\text{NH}_{3})=2.0\times10^{15}-3.8\times10^{16}$~cm$^{-2}$. Now assuming $\text{source size}=0\farcs{5}$ yields $\eta_{\text{BF}}\approx6\times10^{-4}$ and results in the $N(\text{NH}_{3})/\eta_{\text{BF}}$ value used in this work.}}
 \tnote[f]{\tiny{Only one estimate exists in the literature, for the circumbinary envelope scales from \citet{Peng2010} for $\text{beam size}=13\farcs{5}$, $N(\text{HCl})=4.7\times10^{13}$~cm$^{-2}$; assuming $\text{source size}=0\farcs{5}$ yields $\eta_{\text{BF}}=1/730$ and results in the $N(\text{HCl})/\eta_{\text{BF}}$ value used in this work.}}
 }{
 \hline
 Species                   & Name                   & N (cm$^{-2}$)\tmark[b] \& Assumptions & Reference                                        & $T_{\text{ex}}$ (K)\T\B\\
 \hline
 H$_{2}$O                  & Water                  & source A est.                         & \citet{Persson2013}\tmark[c]                     & \T\\
                           &                        & $3.3\times10^{21}$                    &                                                  & $124\pm12$\\
 O$_{2}$                   & Molecular oxygen       & assuming tent. detection              & \citet{Taquet2018}                               & \\
                           &                        & $<2.0\times10^{20}$                   &                                                  & $300$\\
 CO                        & Carbon monoxide        & $1.0\times10^{20}$                    & this work (Appendix~\ref{columndens_IRAS16293B}) & $100-150$\\
 CH$_{3}$OH                & Methanol               & $1.0\times10^{19}$                    & \citet{Jorgensen2016, Jorgensen2018}\tmark[d]    & $300$\\
 H$_{2}$CO                 & Formaldehyde           & $1.9\times10^{18}$                    & \citet{Persson2018}                              & $106\pm13$\\
 C$_{2}$H$_{5}$OH          & Ethanol                & $2.3\times10^{17}$                    & \citet{Jorgensen2018}                            & $300$\\
 CH$_{3}$OCH$_{3}$         & Dimethyl ether         & $2.4\times10^{17}$                    & \citet{Jorgensen2018}                            & $125$\\
 HCOOCH$_{3}$              & Methyl formate         & $2.6\times10^{17}$                    & \citet{Jorgensen2018}                            & $300$\\
 CH$_{2}$OHCHO             & Glycolaldehyde         & $3.2\times10^{16}$                    & \citet{Jorgensen2016}\tmark[d]                   & $300$\\
 CH$_{3}$COOH              & Acetic acid            & $2.8\times10^{15}$                    & \citet{Jorgensen2016}\tmark[d]                   & $300$\\
 CH$_{3}$CHO               & Acetaldehyde           & $1.2\times10^{17}$                    & \citet{Jorgensen2018}                            & $125$\\
 c-C$_{2}$H$_{4}$O         & Ethylene oxide         & $5.4\times10^{15}$                    & \citet{Lykke2017}                                & $125$\\
 CH$_{2}$CHOH              & Vinyl alcohol          & $<1.8\times10^{15}$                   & \citet{Lykke2017}                                & $125$\\
 HCOOH                     & Formic acid            & $5.6\times10^{16}$                    & \citet{Jorgensen2018}                            & $300$\\
 aGg\textquotesingle-((CH$_{2}$OH)$_{2}$) & aGg\textquotesingle-Ethylene glycol & $5.2\times10^{16}$                    & \citet{Jorgensen2016}\tmark[d]                & $300$\\
 gGg\textquotesingle-((CH$_{2}$OH)$_{2}$) & gGg\textquotesingle-Ethylene glycol & $4.7\times10^{16}$                    & \citet{Jorgensen2016}\tmark[d]                & $300$\\
 CH$_{3}$OCH$_{2}$OH       & Methoxymethanol        & $1.4\times10^{17}$                    & Manigand et al. subm.                            & $130$\\
 C$_{2}$H$_{5}$CHO         & Propanal               & $2.2\times10^{15}$                    & \citet{Lykke2017}                                & $125$\\
 (CH$_{3}$)$_{2}$CO        & Acetone                & $1.7\times10^{16}$                    & \citet{Lykke2017}                                & $125$\\
 NH$_{2}$CHO               & Formamide              & $9.5\times10^{15}$                    & \citet{Coutens2016}                              & $300$\\
 NH$_{3}$                  & Ammonia                & circumbinary envelope est.            & \citet{Hily-Blant2010},                          & \\
                           &                        & $<6.1\times10^{19}$                   & \citet{vanDishoeck1995},                         & $8-30$\\
                           &                        &                                       & \citet{Mundy1990},                               & \\
                           &                        &                                       & \citet{Mizuno1990}\tmark[e]                      & \\											
 HCN                       & Hydrogen cyanide       & $5.0\times10^{16}$                    & this work (Appendix~\ref{columndens_IRAS16293B}) & $120$\\
 HNC                       & Hydrogen isocyanide    & $<5.0\times10^{16}$                   & this work (Appendix~\ref{columndens_IRAS16293B}) & $120$\\
 CH$_{3}$CN                & Methyl cyanide         & $4.0\times10^{16}$                    & \citet{Calcutt2018a}                             & $110\pm10$\\
 CH$_{3}$NC                & Methyl isocyanide      & $2.0\times10^{14}$                    & \citet{Calcutt2018b}                             & $150\pm20$\\
 HNCO                      & Isocyanic acid         & $3.7\times10^{16}$                    & \citet{Ligterink2017}                            & $100$\\
 HOCN                      & Cyanic acid            & $<3.0\times10^{13}$                   & \citet{Ligterink2017}                            & $100$\\
 HC$_{3}$N                 & Cyanoacetylene         & $1.8\times10^{14}$                    & \citet{Calcutt2018a}                             & $100\pm20$\\
 H$_{2}$S                  & Hydrogen sulphide      & $1.7\times10^{17}$                    & \citet{Drozdovskaya2018}                         & $125$\\
 OCS                       & Carbonyl suplhide      & $2.5\times10^{17}$                    & \citet{Drozdovskaya2018}                         & $125$\\
 CH$_{3}$SH                & Methyl mercaptan       & $4.8\times10^{15}$                    & \citet{Drozdovskaya2018}                         & $125$\\
 CS                        & Carbon monosulphide    & $3.9\times10^{15}$                    & \citet{Drozdovskaya2018}                         & $125$\\
 H$_{2}$CS                 & Thioformaldehyde       & $1.3\times10^{15}$                    & \citet{Drozdovskaya2018}                         & $125$\\
 S$_{2}$                   & Disulphur              & $<1.9\times10^{16}$                   & \citet{Drozdovskaya2018}                         & $125$\\
 SO$_{2}$                  & Sulphur dioxide        & $1.3\times10^{15}$                    & \citet{Drozdovskaya2018}                         & $125$\\
 SO                        & Sulphur monoxide       & $4.4\times10^{14}$                    & \citet{Drozdovskaya2018}                         & $125$\\
 C$_{2}$H$_{5}$SH          & Ethyl mercaptan        & $<3.2\times10^{15}$                   & \citet{Drozdovskaya2018}                         & $125$\\
 H$_{2}$S$_{2}$            & Disulphane             & $<7.9\times10^{14}$                   & \citet{Drozdovskaya2018}                         & $125$\\
 HS$_{2}$                  & Disulphanide           & $<4.4\times10^{14}$                   & \citet{Drozdovskaya2018}                         & $125$\\
 PO                        & Phosphorus monoxide    & $<4.4\times10^{14}$                   & this work (Appendix~\ref{columndens_IRAS16293B}) & $125$\\
 PN                        & Phosphorus mononitride & $<2.1\times10^{13}$                   & this work (Appendix~\ref{columndens_IRAS16293B}) & $125$\\
 HCl                       & Hydrogen chloride      & circumbinary envelope est.            & \citet{Peng2010}\tmark[f]                        & \\
                           &                        & $<3.4\times10^{16}$                   &                                                  & $80$\\
 CH$_{3}$Cl                & Methyl chloride        & $4.6\times10^{14}$                    & \citet{Fayolle2017}                              & $102\pm3$\\
 NH$_{2}$CH$_{2}$COOH      & Glycine                & $<9.2\times10^{14}$                   & this work (Appendix~\ref{columndens_IRAS16293B}) & $300$\B\\
\hline}
\clearpage

\ctable[
 caption = {Bulk quantities of volatiles in 67P/C--G as measured with ROSINA\tmark.},
 label = {tbl:abunvalues_67P}
 ]{@{\extracolsep{\fill}}lll}{
\tnote{All bulk abundances are from \citealt{Rubin2019a}, unless indicated otherwise (see also Altwegg et al. in press).}
\tnote[b]{Schuhmann et al. under rev.}
\tnote[c]{\citet{Calmonte2016}}
\tnote[d]{Rivilla et al. in prep.} 
\tnote[e]{\citet{Dhooghe2017}}
\tnote[f]{\citet{Fayolle2017}}
\tnote[g]{Hadraoui et al. (in press) with error bars based on the range of glycine abundances reported in \citet{Altwegg2016}}
 }{
 \hline
 Species                                                                         & Name                                               & Abundance rel. to H$_{2}$O (\%)\T\B\\
 \hline
 H$_{2}$O                                                                        & Water                                              & $100$\T\\
 O$_{2}$                                                                         & Molecular oxygen                                   & $3.1\pm1.1$\\
 CO                                                                              & Carbon monoxide                                    & $3.1\pm0.9$\\
 CH$_{3}$OH                                                                      & Methanol                                           & $0.21\pm0.06$\tmark[b]\\
 H$_{2}$CO                                                                       & Formaldehyde                                       & $0.32\pm0.10$\tmark[b]\\
 C$_{2}$H$_{5}$OH $+$ CH$_{3}$OCH$_{3}$                                          & Ethanol $+$ Dimethyl ether                         & $0.039\pm0.023$\tmark[b]\\
 HCOOCH$_{3}$ $+$ CH$_{2}$OHCHO $+$ CH$_{3}$COOH                                 & Methyl formate $+$ Glycolaldehyde $+$ Acetic acid  & $0.0034\pm0.0020$\tmark[b]\\
 CH$_{3}$CHO $+$ c-C$_{2}$H$_{4}$O $+$ CH$_{2}$CHOH                              & Acetaldehyde $+$ Ethylene oxide $+$ Vinyl alcohol  & $0.047\pm0.017$\tmark[b]\\
 HCOOH                                                                           & Formic acid                                        & $0.013\pm0.008$\tmark[b]\\
 aGg\textquotesingle-((CH$_{2}$OH)$_{2}$) $+$ gGg\textquotesingle-((CH$_{2}$OH)$_{2}$) $+$ CH$_{3}$OCH$_{2}$OH & aGg\textquotesingle- and gGg\textquotesingle-Ethylene glycol $+$ Methoxymethanol    & $0.011\pm0.007$\tmark[b]\\
 C$_{2}$H$_{5}$CHO $+$ (CH$_{3}$)$_{2}$CO $+$ CH$_{3}$CHCH$_{2}$O                & Propanal $+$ Acetone $+$ Propylene oxide           & $0.0047\pm0.0024$\tmark[b]\\
 NH$_{2}$CHO                                                                     & Formamide                                          & $0.0040\pm0.0023$\\
 NH$_{3}$                                                                        & Ammonia                                            & $0.67\pm0.20$\\
 HCN $+$ HNC                                                                     & Hydrogen cyanide $+$ Hydrogen isocyanide           & $0.14\pm0.04$\\
 CH$_{3}$CN $+$ CH$_{3}$NC                                                       & Methyl cyanide $+$ Methyl isocyanide               & $0.0059\pm0.0034$\\
 HNCO $+$ HOCN                                                                   & Isocyanic acid $+$ Cyanic acid                     & $0.027\pm0.016$\\
 HC$_{3}$N $+$ HC$_{2}$NC                                                        & Cyanoacetylene $+$ Isocyanoacetylene               & $0.00040\pm0.00023$\\
 H$_{2}$S                                                                        & Hydrogen sulphide                                   & $1.10\pm0.46$\tmark[c]\\
 OCS                                                                             & Carbonyl sulphide                                   & $0.041^{+0.082}_{-0.020}$\tmark[c]\\
 CH$_{3}$SH                                                                      & Methyl mercaptan                                   & $0.038^{+0.079}_{-0.028}$\tmark[c]\\
 H$_{2}$CS                                                                       & Thioformaldehyde                                   & $0.0027^{+0.0058}_{-0.0024}$\tmark[c]\\
 CS$_{2}$                                                                        & Carbon disulphide                                  & $0.0057^{+0.0114}_{-0.0028}$\tmark[c]\\
 S$_{2}$                                                                         & Disulphur                                          & $0.0020^{+0.0040}_{-0.0010}$\tmark[c]\\
 SO$_{2}$                                                                        & Sulphur dioxide                                    & $0.127^{+0.254}_{-0.064}$\tmark[c]\\
 SO                                                                              & Sulphur monoxide                                   & $0.071^{+0.142}_{-0.037}$\tmark[c]\\
 C$_{2}$H$_{5}$SH $+$ (CH$_{3}$)$_{2}$S                                          & Ethyl mercaptan $+$ Dimethyl sulphide              & $0.00058^{+0.00123}_{-0.00049}$\tmark[c]\\
 H$_{2}$S$_{2}$                                                                  & Disulphane                                         & $\leq0.0006042^{+0.005778}_{-0.0005778}$\tmark[c]\\
 HS$_{2}$                                                                        & Disulphanide                                       & $\leq0.000106^{+0.000954}_{-0.0000954}$\tmark[c]\\
 PO                                                                              & Phosphorus monoxide                                & $0.011^{+0.022}_{-0.0006}$\tmark[d]\\
 PN                                                                              & Phosphorus mononitride                             & $<0.0011^{+0.0022}_{-0.00006}$\tmark[d]\\
 HCl                                                                             & Hydrogen chloride                                  & $0.014^{+0.045}_{-0.012}$\tmark[e]\\
 CH$_{3}$Cl                                                                      & Methyl chloride                                    & $0.000056^{+0.000298}_{-0.000052}$\tmark[f]\\
 NH$_{2}$CH$_{2}$COOH                                                            & Glycine                                            & $0.000017^{+0.249983}_{-0.000017}$\tmark[g]\B\\
\hline}
\clearpage

\ctable[
 caption = {Correlation coefficients between volatiles towards IRAS~16293-2422~B on disc-scales and in the bulk of 67P/C--G.},
 label = {tbl:corr_coeff}
 ]{@{\extracolsep{\fill}}llll}{
 }{
 \hline
                                                          &                                             & Linear scaling     & Logarithmic scaling \T\B\\
 \hline
 CHO-bearing molecules                                    &                                             &                    & \T\B\\
 \hline
                                                          & Pearson correlation coefficient ($r$)       & $1.0$              & $0.95$\T\\
	                                                        & Spearman's correlation coefficient ($\rho$) & $0.88$             & $0.88$\\
											                                    & Spearman's significance (2-tailed)          & $8.1\times10^{-5}$ & $8.1\times10^{-5}$\\
											                                    & Sample size                                 & $13$               & $13$\B\\
 \hline
 CHO-bearing molecules (without CO and H$_{2}$O)          &                                             &                    & \T\B\\
 \hline
                                                          & Pearson correlation coefficient ($r$)       & $1.0$              & $0.91$\T\\
	                                                        & Spearman's correlation coefficient ($\rho$) & $0.80$             & $0.80$\\
											                                    & Spearman's significance (2-tailed)          & $3.1\times10^{-3}$ & $3.1\times10^{-3}$\\
										                                   	  & Sample size                                 & $11$               & $11$\B\\
 \hline
 CHO-bearing molecules (without CO, H$_{2}$O and O$_{2}$) &                                             &                    & \T\B\\
 \hline
                                                          & Pearson correlation coefficient ($r$)       & $0.61$             & $0.88$\T\\
	                                                        & Spearman's correlation coefficient ($\rho$) & $0.73$             & $0.73$\\
											                                    & Spearman's significance (2-tailed)          & $1.5\times10^{-2}$ & $1.5\times10^{-2}$\\
											                                    & Sample size                                 & $10$               & $10$\B\\
 \hline
 N-bearing molecules                                      &                                             &                    & \T\B\\
 \hline
                                                          & Pearson correlation coefficient ($r$)       & $0.98$             & $0.86$\T\\
	                                                        & Spearman's correlation coefficient ($\rho$) & $0.93$             & $0.93$\\
											                                    & Spearman's significance (2-tailed)          & $2.5\times10^{-3}$ & $2.5\times10^{-3}$\\
											                                    & Sample size                                 & $7$                & $7$\B\\
 \hline
 S-bearing molecules                                      &                                             &                    & \T\B\\
 \hline
                                                          & Pearson correlation coefficient ($r$)       & $0.50$             & $0.49$\T\\
	                                                        & Spearman's correlation coefficient ($\rho$) & $0.32$             & $0.32$\\
											                                    & Spearman's significance (2-tailed)          & $3.5\times10^{-1}$ & $3.5\times10^{-1}$\\
											                                    & Sample size                                 & $11$               & $11$\B\\
 \hline
 P- and Cl-bearing molecules                              &                                             &                    & \T\B\\
 \hline
                                                          & Pearson correlation coefficient ($r$)       & $0.71$             & $0.44$\T\\
	                                                        & Spearman's correlation coefficient ($\rho$) & $0.40$             & $0.40$\\
											                                    & Spearman's significance (2-tailed)          & $6.0\times10^{-1}$ & $6.0\times10^{-1}$\\
											                                    & Sample size                                 & $4$                & $4$\B\\
\hline}
\clearpage

%%%%%%%%%%%%%%%%%%%%%%%%%%%%%%%%%%%%%%%%%%%%%%%%%%%%%%%%%%%%%%%%%%%%%%%%%%%%%%%
\section{Results}
\label{results}

\begin{figure}
 \centering
  \includegraphics[width=0.45\textwidth,height=0.8\textheight,keepaspectratio]{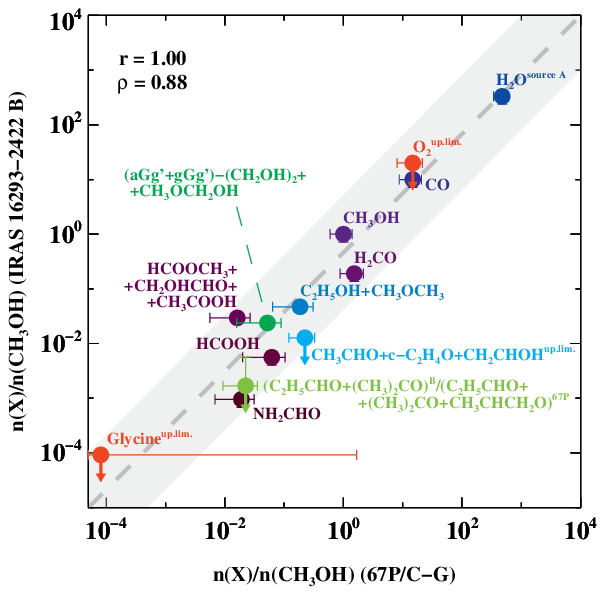}
 \caption{The abundance of CHO-bearing molecules relative to methanol detected towards the one-beam offset position from IRAS~16293-2422~B versus that measured in 67P/C--G. Each molecule is marked with a unique color. The shaded region corresponds to an order of magnitude scatter about the linear correlation. The Pearson ($r$) and Spearman ($\rho$) correlation coefficients are given in the upper left corner. ``up.lim.'' indicates the values that are protostellar upper limits; and ``Source A'' indicates the value that is an estimate based on IRAS~16293-2422~A.}
 \label{fgr:corrplotCHO}
\end{figure}

\begin{figure}
 \centering
  \includegraphics[width=0.45\textwidth,height=0.8\textheight,keepaspectratio]{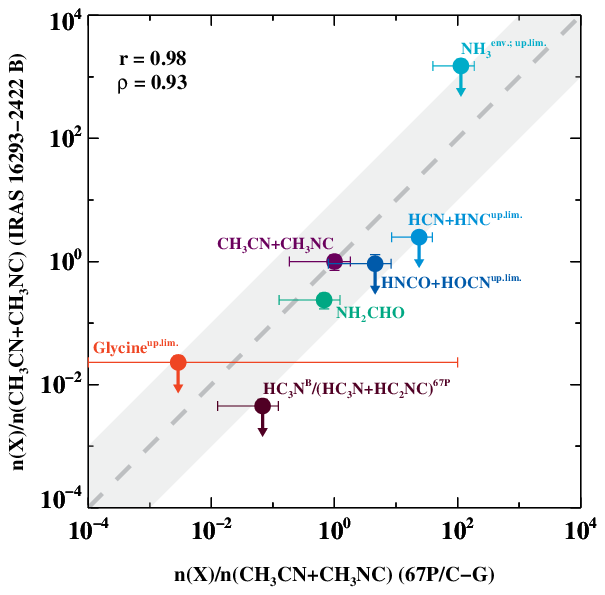}
 \caption{The abundance of N-bearing molecules relative to methyl cyanide detected towards the one-beam offset position from IRAS~16293-2422~B versus that measured in 67P/C--G. Each molecule is marked with a unique color. The shaded region corresponds to an order of magnitude scatter about the linear correlation. The Pearson ($r$) and Spearman's ($\rho$) correlation coefficients are given in the upper left corner. ``up.lim.'' indicates the values that are protostellar upper limits; and ``env.'' indicates the value that is an estimate based on the circumbinary envelope.}
 \label{fgr:corrplotN}
\end{figure}

\begin{figure}
 \centering
  \includegraphics[width=0.45\textwidth,height=0.8\textheight,keepaspectratio]{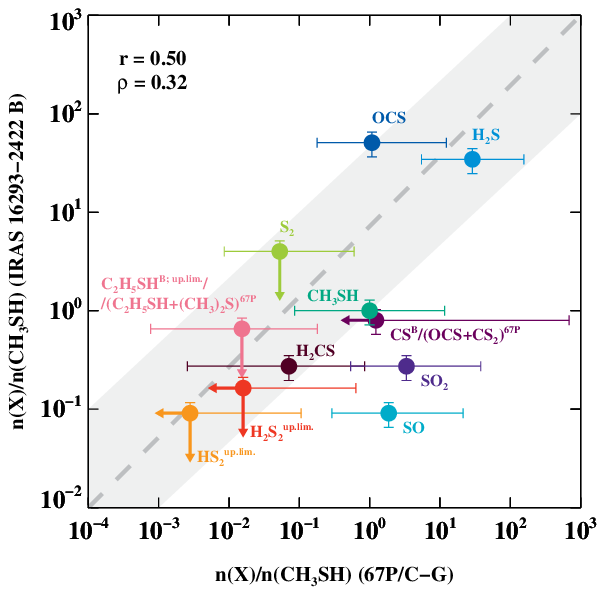}
 \caption{The abundance of S-bearing molecules relative to methyl mercaptan detected towards the one-beam offset position from IRAS~16293-2422~B versus that measured in 67P/C--G. Each molecule is marked with a unique color. The shaded region corresponds to an order of magnitude scatter about the linear correlation. The Pearson ($r$) and Spearman's ($\rho$) correlation coefficients are given in the upper left corner. ``up.lim.'' indicates the values that are protostellar upper limits.}
 \label{fgr:corrplotS}
\end{figure}

\begin{figure}
 \centering
  \includegraphics[width=0.45\textwidth,height=0.8\textheight,keepaspectratio]{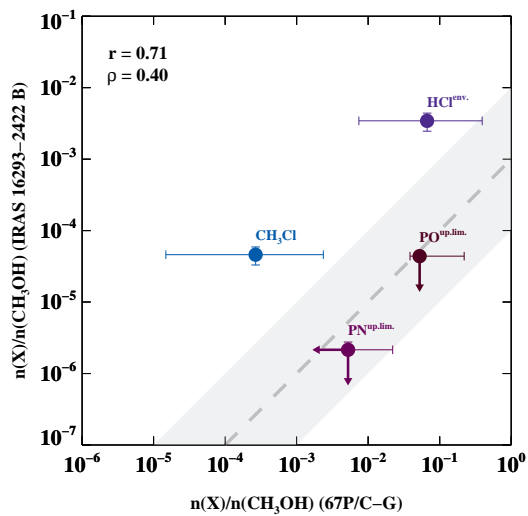}
 \caption{The abundance of P- and Cl-bearing molecules relative to methanol detected towards the one-beam offset position from IRAS~16293-2422~B versus that measured in 67P/C--G. Each molecule is marked with a unique color. The shaded region corresponds to an order of magnitude scatter about the linear correlation. The Pearson ($r$) and Spearman's ($\rho$) correlation coefficients are given in the upper left corner. ``up.lim.'' indicates the values that are protostellar upper limits; and ``env.'' indicates the value that is an estimate based on the circumbinary envelope.}
 \label{fgr:corrplotother}
\end{figure}

\subsection{Correlations between IRAS~16293-2422~B and 67P/Churyumov-Gerasimenko}
Figs.~\ref{fgr:corrplotCHO},~\ref{fgr:corrplotN},~\ref{fgr:corrplotS}, and~\ref{fgr:corrplotother} show log-log plots of the observed relative abundances of 67P/C--G and IRAS~16293-2422~B used to search for correlations between bulk cometary volatiles and protoplanetary disc-materials. The volatiles have been partitioned into chemical families based on the elements that they carry. The reference species for computing the relative abundances differ per chemical family and have been chosen based on the interstellar chemical pathways to form such chemically related molecules. The majority of cometary data points represent bulk abundances as derived from ROSINA measurements, as explained in Section~\ref{67P} and published in \citet{Rubin2019a}. The protostellar values represent the material entering the protoplanetary disc around source B that have been derived for a region $0\farcs{5}$ in size and stem from the series of papers from the PILS team at a relative accuracy of $10-20$ per cent. Variations in the $T_{\text{ex}}$ of $\sim20$ per cent ($25-60$~K for $T_{\text{ex}}=125-300$~K) change the derived column densities by $<10$ per cent \citep{Jorgensen2018}. Some previously unpublished and newly derived column densities (or at least upper limits) that are used for this study are presented in Appendix~\ref{columndens_IRAS16293B}. For molecules for which PILS data are insufficient, either estimates or ranges are provided based on earlier observations. All the protostellar and cometary abundances used for the analysis are tabulated in Tables~\ref{tbl:abunvalues_B} and~\ref{tbl:abunvalues_67P}, respectively. The error bars have been computed by considering [minimum, maximum] ratio ranges for the case of asymmetric errors, and via error propagation equations of normally distributed values for the case of symmetric errors. The full inventory of species detected towards IRAS~16293-2422 is provided in Appendix~\ref{molecules_IRAS16293}.

Fig.~\ref{fgr:corrplotCHO} displays the abundance of CHO-bearing molecules relative to methanol (CH$_{3}$OH) detected towards the one-beam offset position from source B versus that measured in 67P/C--G. The reference species selected is CH$_{3}$OH for the CHO-bearing family, because it is thought to form on grain surfaces via sequential hydrogenation of CO \citep{TielensHagen1982, Fuchs2009} and to be a key precursor to the synthesis of larger O-bearing complex organic molecules \citep{Garrod2008, Fedoseev2017}. The ROSINA mass spectrometer cannot unambiguously distinguish isomers as they have the same mass; hence, the measurements at mass $44$~u/e are a combination of acetaldehyde (CH$_{3}$CHO), ethylene oxide (c-C$_{2}$H$_{4}$O), and vinyl alcohol (CH$_{2}$CHOH); at mass $46$~u/e are a combination of ethanol (C$_{2}$H$_{5}$OH) and dimethyl ether (CH$_{3}$OCH$_{3}$); at mass $60$~u/e are a combination of methyl formate (HCOOCH$_{3}$), glycolaldehyde (CH$_{2}$OHCHO), and acetic acid (CH$_{3}$COOH); and at mass $62$~u/e are a combination of both ethylene glycol forms ((CH$_{2}$OH)$_{2}$) and methoxymethanol (CH$_{3}$OCH$_{2}$OH). Spectroscopic observations have the ability to distinguish isomers; however, the observed protostellar abundances have been summed to make an appropriate comparison with cometary measurements. The cometary value for glycine does not stem from the same period as bulk abundances for the majority of the other species analysed in this work, as continuous data were not available for this molecule during the mission. The value used stems from dedicated models of glycine in 67P/C--G, which inferred it to be desorbing from the nucleus and from icy mantles of dust particles ejected from the nucleus into the coma (i.e., a distributed source, \citealt{Altwegg2016}), while being mixed with water ice in both of these sources (Hadraoui et al. in press). The error bars are based on the range of measured glycine abundances reported in \citet{Altwegg2016}. The protostellar abundances of O$_{2}$, CH$_{2}$CHOH and glycine are upper limits, which means that these points may shift lower along the ordinate. The protostellar abundance of H$_{2}$O is an estimate based on the value derived for source A (see Table~\ref{tbl:abunvalues_B} for details). The C$_{2}$H$_{5}$CHO data point is a currently best-possible estimate, as the ROSINA measurement at mass $58$~u/e is a combination of propanal (C$_{2}$H$_{5}$CHO), acetone ((CH$_{3}$)$_{2}$CO), and propylene oxide (CH$_{3}$CHCH$_{2}$O), while the column density of the latter is currently unavailable for IRAS~16293-2422. The figure appears to display a linear correlation between the two sets of abundances (including those that are upper limits) with a Pearson correlation coefficient of $1.00$ and a Spearman's correlation coefficient of $0.88$ (at a two-tailed significance of $8.1\times10^{-5}$; Table~\ref{tbl:corr_coeff}), which implies that cometary and protostellar CHO-bearing volatiles are related.

Fig.~\ref{fgr:corrplotN} displays the abundance of N-bearing molecules relative to methyl cyanide (CH$_{3}$CN) detected towards the one-beam offset position from source B versus that measured in 67P/C--G. The ROSINA mass spectrometer measurement at mass $27$~u/e is a combination of hydrogen isocyanide (HNC) and hydrogen cyanide (HCN); at mass $41$~u/e is a combination of methyl isocyanide (CH$_{3}$NC) and methyl cyanide; at mass $43$~u/e is a combination of isocyanic acid (HNCO) and cyanic acid (HOCN); the protostellar values have been summed for these pairs of molecules accordingly. The cometary value for glycine is the best available, as discussed in the above paragraph. The protostellar abundances of HNC, HOCN and glycine are upper limits, which means that these points may shift lower along the ordinate. The HC$_{3}$N data point is a currently best-possible estimate, as the ROSINA measurement at mass $51$~u/e is a combination of cyanoacetylene (HC$_{3}$N) and isocyanoacetylene (HC$_{2}$NC), while the column density of the latter is currently unavailable for IRAS~16293-2422. The protostellar abundance of NH$_{3}$ is an estimated upper limit based on circumbinary envelope-scale observations (see Table~\ref{tbl:abunvalues_B} for details). This figure appears to display a linear correlation between the two sets of abundances  (including those that are upper limits) with a Pearson correlation coefficient of $0.98$ and a Spearman's correlation coefficient of $0.93$ (at a two-tailed significance of $2.5\times10^{-3}$; Table~\ref{tbl:corr_coeff}), which implies that cometary and protostellar N-bearing volatiles are related. The reference species selected is CH$_{3}$CN for the N-bearing family, whose origin is still unclear. It can be formed on grain surfaces with some contributions from gas-phase reactions \citep{Calcutt2018a}. CH$_{3}$CN is the analogous 6-atom molecule with a methyl (CH$_{3}$-) functional group to CH$_{3}$OH. CH$_{3}$NC had to be added to the chosen reference species CH$_{3}$CN due to their identical mass being indistinguishable for the ROSINA instrument; however, the protostellar column density of CH$_{3}$NC is two orders of magnitude lower than that of CH$_{3}$CN, and hence, likely only makes a minor difference.

Fig.~\ref{fgr:corrplotS} displays the abundance of S-bearing molecules relative to methyl mercaptan (CH$_{3}$SH) detected towards the one-beam offset position from source B versus that measured in 67P/C--G. In ROSINA data, the CS peak suffers from interference with the strong signal from abundant CO$_{2}$ at mass $44$~u/e. Over a short fly-by in March 2015, when CO$_{2}$ was scarcer, it was derived that all CS detected by ROSINA can be explained by the fragmentation of primarily CS$_{2}$ and secondarily, of OCS \citep{Calmonte2016}. Therefore, the amount of CS as a radical in cometary ice is small, if existent at all. Summing the abundances of CS$_{2}$ and OCS ices yields an upper limit for the amount of CS radicals incorporated into 67P/C--G at some point since the prestellar phase, assuming that all CS radicals are converted into either OCS or CS$_{2}$ by the time they become constituents of cometary ice. This is the CS upper limit used in Fig.~\ref{fgr:corrplotS}. Cometary values for H$_{2}$CS, CH$_{3}$SH and C$_{2}$H$_{5}$SH do not stem from the same period as bulk abundances for the majority of the other species analysed in this work, as continuous data were not available for these two molecules during the mission. Hence, these two numbers may be somewhat less representative of the bulk, but are the best available (see \citealt{Calmonte2016} for details). The cometary abundances of HS$_{2}$ and H$_{2}$S$_{2}$ are upper limits, which means that these points may shift to the left along the abscissa. The protostellar abundances of S$_{2}$, HS$_{2}$, H$_{2}$S$_{2}$ and C$_{2}$H$_{5}$SH are upper limits, which means that these points may shift lower along the ordinate. The C$_{2}$H$_{5}$SH data point is a currently best-possible estimate, as the ROSINA measurement at this mass of $62$~u/e is a combination of ethyl mercaptan (C$_{2}$H$_{5}$SH) and dimethyl sulphide ((CH$_{3}$)$_{2}$S), while spectroscopy of dimethyl sulphide is not yet available, which inhibits its search in the ALMA data. The figure appears to display a linear correlation between the two sets of abundances  (including those that are upper limits) with a Pearson correlation coefficient of $0.50$ and a Spearman's correlation coefficient of $0.32$ (at a two-tailed significance of $0.35$; Table~\ref{tbl:corr_coeff}), which implies that cometary and protostellar S-bearing volatiles are related. These $r$- and $\rho$-values are lower than that of CHO- and N-bearing species, potentially due to the larger fraction of upper limits and best-effort estimates used in the S-bearing family. Furthermore, S-bearing molecules span a smaller range of relative abundances than the CHO- and N-bearing species. The reference species selected is CH$_{3}$SH for the S-bearing family, which is formed on grain surfaces from atomic sulphur and CS hydrogenations \citep{Vidal2017, Lamberts2018}. CH$_{3}$SH is the analogous 6-atom molecule with a methyl (CH$_{3}$-) functional group to CH$_{3}$OH and CH$_{3}$CN.

Fig.~\ref{fgr:corrplotother} displays the abundance of P- and Cl-bearing molecules relative to methanol detected towards the one-beam offset position from source B versus that measured in 67P/C--G. The cometary abundance of PN is an upper limit, as its mass peak overlaps with those of CHS and $^{13}$CS, and suffers from strong interference with the peak of $^{13}$CO$_{2}$ (Rivilla et al. in prep.). The cometary values for PN, PO and CH$_{3}$Cl do not stem from the same period as bulk abundances for the majority of the other species analysed in this work, as continuous data were not available for these molecules during the mission. Hence, as for H$_{2}$CS, CH$_{3}$SH and C$_{2}$H$_{5}$SH, these numbers may be somewhat less representative of the bulk, but are the best available (see Rivilla et al. in prep. and \citealt{Fayolle2017} for details). The protostellar abundances for PO and PN are the currently best-available upper limit estimates, which will soon be tested with new dedicated ALMA observations (project-id: 2018.1.01496, PI: V{\'i}ctor M. Rivilla). The protostellar abundance of HCl is an estimate based on circumbinary envelope-scale observations (see Table~\ref{tbl:abunvalues_B} for details). The figure appears to display a linear correlation between the two sets of abundances  (including those that are upper limits) with a Pearson correlation coefficient of $0.71$ and a Spearman's correlation coefficient of $0.40$ (at a two-tailed significance of $0.6$; Table~\ref{tbl:corr_coeff}), although more data points are desirable. The linear correlation is not one-to-one, but is offset, which may be a result of CH$_{3}$OH being chosen as the reference species. It is not clear what the best reference species is for these exotic species. The correlation tentatively suggests that cometary and protostellar P- and Cl-bearing volatiles are related. These chemical families have never been probed before.

The correlations in Figs.~\ref{fgr:corrplotCHO},~\ref{fgr:corrplotN},~\ref{fgr:corrplotS}, and~\ref{fgr:corrplotother} vary in strength and significance. The relative abundances investigated span a wide range; hence, the correlation coefficients have also been computed with logarithmic scaling (the coefficients given in the figures are derived with the plotted linear scaling). Table~\ref{tbl:corr_coeff} summarizes all the correlation coefficients and their significance. It can be seen that logarithmic scaling lowers the strength of the correlations somewhat in terms of the Pearson correlation coefficients; however, they remain statistically significant. For the case of CHO-bearing molecules, that span the largest range of relative abundances, the correlation has also been scrutinized upon exclusion of simpler species, specifically H$_{2}$O, CO and O$_{2}$. This results in a significant reduction of the Pearson and Spearman's correlation coefficients to $0.61$ and $0.73$ (at a two-tailed significance of $0.015$), respectively. This supports there being significantly more complex organic CHO-bearing species in the comet than towards the protostar.

Beyond the chemical relevance of the three reference species (CH$_{3}$OH, CH$_{3}$CN, and CH$_{3}$SH) for their respective chemical families, they are also expected to be present predominantly on small scales in the hot inner regions around protostars due to their high desorption energies ($5534$, $4680$, $4000$~K, respectively). This consequently makes these molecules more relevant for tracing disc-materials, rather than those that may be thermally desorbed already at lower temperatures, which are easily attained on larger envelope-scales in the system (as is the case for HCN, for example). The reference species CH$_{3}$OH, CH$_{3}$CN, and CH$_{3}$SH ensure one-to-one correlations in Figs.~\ref{fgr:corrplotCHO},~\ref{fgr:corrplotN}, and~\ref{fgr:corrplotS} for the CHO-, N-, and S-bearing families, respectively. A choice of a common reference species (such as CH$_{3}$OH or H$_{2}$O) for all molecules preserves the linear correlations for the members of a single chemical family, but introduces a scaling factor to the linear correlation, i.e., it is no longer one-to-one. This can be seen by comparing Fig.~\ref{fgr:corrplotmerged} with Figs.~\ref{fgr:corrplotmergedCH3OH} and~\ref{fgr:corrplotmergedH2O}. This is analogous to the offset seen for P-and Cl-bearing molecules when using methanol as the reference species (Fig.~\ref{fgr:corrplotother}).

\subsection{Caveats}
\subsubsection{Missed reservoirs}
Not all major reservoirs have been probed with the observations discussed in this work. As discussed in the previous Section, it is not possible to probe individual isomers unambiguously with ROSINA measurements for 67P/C--G, as well as the interfering CS and $^{13}$CO$_{2}$. Meanwhile, with ALMA data at radio frequencies that are sensitive to rotational lines of molecules, it is not possible to determine the abundance of symmetric molecules such as N$_{2}$, CH$_{4}$, CS$_{2}$, and CO$_{2}$; and of atoms such as S. Only non-trivial combinations of data from different instruments can tackle these missed reservoirs of volatiles.

\subsubsection{Representability of the targets}
There is no evidence to suggest that 67P/C--G is in any way an atypical comet. Its bilobate shape is similar to that of comet 103P/Hartley 2 \citep{AHearn2011b} among others, and the trans-Neptunian object (486958)~2014~MU$_{69}$ \citep{Stern2019}. The topographically heterogeneous surface of 67P/C--G dominated by smooth-floored pits appears to be most similar to 81P/Wild 2 \citep{Birch2017}. Its dominant volatile has been shown to be water ice that is hidden in the interior and almost completely absent from the surface, which is again typical of all the comets studied thus far \citep{Filacchione2016a}. The low-density and high porosity ($72-74$ per cent) of the nucleus of 67P/C--G is comparable to that of comet 9P/Tempel 1 \citep{Patzold2016}. The coma has been quite tenuous in comparison to the brightest comets that have been observed, such as Hale--Bopp whose production rates were roughly two orders of magnitude higher (Altwegg et al. in press). The chemical richness observed on 67P/C--G is very likely a mere consequence of the superior measurement techniques (long-term monitoring at close distances coupled with high sensitivity of \textit{Rosetta}'s scientific payload). This is supported by the detection of ethylene glycol and formamide on comets C/2012~F6~(Lemmon) and C/2013~R1~(Lovejoy), as well as ethanol and glycolaldehyde in the latter target \citep{Biver2014,Biver2015b}. The determined volatile composition does not show any major differences from that seen in other comets (as reviewed in \citealt{Cochran2015, DelloRusso2016}).

Likewise, there is no firm support for IRAS~16293-2422 being in any way a unique young stellar object in terms of the chemical abundances and diversity that is observed \citep{Jorgensen2004, Taquet2015}. The short distance to this source facilitates the detection of all the minor and weakly-emitting molecules. For example, complex organic molecule abundances towards L483 as observed with ALMA compare well to those of IRAS~16293-2422~B \citep{Jacobsen2018b}. From the point of view of the physical structure, there also does not seem to be anything out of the ordinary within the large morphological diversity that is seen in star-forming regions. Multiplicity appears to be common for Class 0 and I sources \citep{Tobin2016}. Source A has also been suggested to be binary in itself (e.g., \citealt{Hernandez-Gomez2019b}). The deuteration of water as measured via the HDO/H$_{2}$O ratio in IRAS~16293-2422 is in range of other deeply embedded low-mass sources on the same spatial scales (fig.~6 of \citealt{Persson2014}), hence, suggesting no drastic temperature differences at the time of water molecule formation in such systems. Further work remains to be done for a larger sample of isolated protostars and the more classical hot corino sources.

On the other hand, it is thought that binaries that are separated by more than disc-scales will not be significantly impacted by neither the passive (heating of the inner collapsing envelope by the protostellar luminosity) nor active (heating by shocks) heating nor the UV flux of their companion. Such conclusions were reached based on $^{13}$CO observations across samples of low-mass protostars \citep{vanKempen2009, Yildiz2013a, Yildiz2015}. The only parts of low-mass binary systems that will be heated and UV-irradiated on scales of up to $\sim1000$~au are the outflow cavities and the cavity walls. This result even holds for species that are enhanced in abundance by UV (e.g., c-C$_{3}$H$_{2}$; \citealt{Murillo2018c}). In IRAS~16293-2422, source B appears to lie at a projected position that overlaps with the northwest outflow stemming from source A \citep{Kristensen2013, Girart2014, vanderWiel2019}. Unfortunately, the inclination angle of the northwest/southeast outflows of source A with respect to the plane of the sky and with respect to source B or the ``bridge'' remains unknown. However, the emission line profiles near source B do not show any evidence for shocks or outflows impinging onto source B from the outside. Consequently, source A is thought to not affect neither the temperature structure nor the UV field in the vicinity of source B, in contrast to a source such as B1-bW \citep{HiranoLiu2014}. This implies that the binary nature of IRAS~16293-2422 is likely not significant in the context of analyses on disc-scales carried out in this work.

%%%%%%%%%%%%%%%%%%%%%%%%%%%%%%%%%%%%%%%%%%%%%%%%%%%%%%%%%%%%%%%%%%%%%%%%%%%%%%%
\section{Discussion}
\label{discussion}

\subsection{Chemical links between comets and the ISM}
The ALMA data analysed in Section~\ref{results} are sensitive to the gases present $\sim70$~au away from protostar B. It is anticipated that the observed gases represent the volatiles that are being transported into the forming protoplanetary disc, which have recently been thermally desorbed upon approach to the protostar, and are likely still present in the solid phase in colder regions of the disc. Thus, these ALMA data provide a unique view into the chemical composition of planet- and comet-forming materials in the low-mass source IRAS~16293-2422~B. The focal one $0.5''$-beam offset position of this work represents some of the most accurate and systematically derived relative abundances for volatiles in a forming Solar-like disc due to the use of optically thin isotopologues, no beam dilution and a relative accuracy of $10-20$ per cent on the derived column densities.

The ROSINA data used in Section~\ref{results} pertain to in situ measurements of the coma gases of 67P/C--G with the majority of the uncertainties being $\sim30$ per cent. These measurements are unique due to the continuous monitoring carried out by the \textit{Rosetta} mission, which allows bulk abundances to be derived for the first time rather than mere snap shots at an isolated moment in time. Hence, the cometary values analysed in this study are the most representative available for the building blocks of our Solar System.

Figs.~\ref{fgr:corrplotCHO},~\ref{fgr:corrplotN},~\ref{fgr:corrplotS}, and~\ref{fgr:corrplotother} display correlations between CHO-, N-, S-, P-, and Cl-bearing volatiles observed in the protostar IRAS~16293-2422~B and comet 67P/C--G with Pearson correlation coefficients in the $[0.50, 1.0]$ and Spearman's correlation coefficients in the $[0.32,0.93]$ ranges. These correlations suggest that volatiles in all low-mass Solar-like systems may be comparable and that some degree of preservation occurs for volatiles from the protostellar phases into comets. This implies that the composition of planetesimals is set, to some extent, in the youngest embedded phase of star formation. Scatter, up to an order of magnitude, is observed. This is particularly noticeable when exploring smaller ranges of relative abundances, such as the case for S-bearing molecules and CHO-bearing complex organic molecules (i.e., excluding H$_{2}$O, CO and O$_{2}$ in Fig.~\ref{fgr:corrplotCHO}). This may stem from the inclusion of upper limits in the analysis, or may be a natural consequence of the slightly different physical evolution of our Solar System from that of IRAS~16293-2422~B.

Figs.~\ref{fgr:corrplotCHO} and~\ref{fgr:corrplotN} show that the relative abundance ratios of 67P/C--G tend to be higher than those of IRAS~16293-2422 for CHO- and N-bearing species. This may indicate that relative to the reference species of a molecular family, the molecules considered in this work have been destroyed at the position near IRAS~16293-2422~B investigated with ALMA data in this work. Potential destruction may occur through gas-phase chemistry upon thermal desorption. Alternatively, it may be that more of the investigated molecules have been produced by the time of incorporation into the comet. For example, chemical modeling suggests that CO will be converted to CH$_{3}$OH, CO$_{2}$ and hydrocarbons within protoplanetary discs \citep{Bosman2018b}. Higher relative quantities may stem from older material, which has given chemical reactions more time to produce more chemically complex species at elevated dust temperatures and UV fluxes during collapse through grain-surface chemical reactions (e.g., \citealt{Drozdovskaya2014, Drozdovskaya2016}). Finally, earlier works have indicated that the amount of methanol in comets is generally lower than in protostellar regions \citep{Oberg2011c2d}. This could also result in higher relative abundances being seen for 67P/C--G when using CH$_{3}$OH as a reference species (Figs.~\ref{fgr:corrplotCHO} and~\ref{fgr:corrplotmergedCH3OH}). Only a dedicated model combining cometesimal formation, global physical evolution of the star-disc system, and simultaneous chemistry can shed light on these scenarios.

In comparison to earlier work, the strong correlation between Hale--Bopp's and ISM relative abundances of CHO- and N-bearing molecules (fig.~$3a$ and $3b$ of \citealt{Bockelee-Morvan2000}) have now been confirmed for the case of a Jupiter-family comet, 67P/C--G, and for disc-scale rather than cloud-scale materials. Such correlations for the case of S-bearing species have been established for the first time in this work. This connection may have been missed due to the data on Hale--Bopp (fig.~$3c$ of \citealt{Bockelee-Morvan2000}) being a mere snapshot of its coma composition at the time that the observations were carried out. S-bearing species, especially S$_{3}$ and S$_{4}$, are strongly associated with high dust densities in the coma \citep{Calmonte2016}, potentially implying that remote observations may be picking up S-bearing species originating from the nucleus as well as from a distributed source \citep{CottinFray2008, Altwegg2017b}. Contributions from S-bearing species stemming from the refractory dust may mask the correlation in volatiles. Alternatively, the abundances of S-bearing molecules may be more variable than others across star-forming regions. Most notably, SO and SO$_{2}$ are well-known outflow tracers, which vary in brightness on cloud scales. In the work of \citet{Bockelee-Morvan2000}, ISM observations were a compilation of data on the region L1157-B1 shocked by a nearby low-mass protostar and the hot cores associated with forming high-mass protostars W3(H$_{2}$O), G34.3+0.15, and Orion KL (Hot Core and Compact Ridge). Observations of such a diverse set of targets are sensitive to different spatial scales and are likely to probe several different components of star-forming systems simultaneously. The correlations seen in this work have been strengthened in the case of CHO-bearing molecules, in particular for CO and the estimate for H$_{2}$O, most likely due to the probing of identical spatial scales by the ALMA data on IRAS~16293-2422. This may also be the reason for the reduction of deviation from the linear correlation for HNCO and HC$_{3}$N in the N-bearing family.

When exploring the correlations between cometary and ISM molecules, \citet{Bockelee-Morvan2000} used CH$_{3}$OH to scale quantities for CHO-bearing molecules and HCN for N- and S-bearing species. Hence, CH$_{3}$OH was the only chemically relevant scaling factor used. The choice of normalizing by HCN for N- and S-bearing species was justified on the basis of the comparable D/H ratio as measured in CH$_{3}$OH and HCN. Now, 20 years later, it is not so clear whether this is something that holds true for protostellar sources in general. HCN was also chosen due to its high abundance (or production rate) in Hale--Bopp at that time. Beyond the argument of spatial scales of ISM observations (discussed in the above paragraph, and as HCN is expected to thermally desorb on envelope-scales at cool temperatures), the newly uncovered correlation in S-bearing molecules may have emerged thanks to the choice of a more representative reference molecule (that is CH$_{3}$SH).

Formamide can be classified as either a CHO- or an N-bearing family member. However, it appears to lie closer to the linear correlation seen in the N-bearing species, while it tends to be more of an outlier in the CHO-bearing family. This may suggest that it is more strongly chemically related to CH$_{3}$CN and the N-bearing molecules, rather than CH$_{3}$OH and the CHO-bearing species. Formamide has already been suggested to be closely related to HNCO based on observational data \citep{Bisschop2007, Lopez-Sepulcre2015, Coutens2016}. Laboratory data are indicating that formamide is a result of combined NO hydrogenation and photolysis in CO-rich ices, and therefore linked to the formation of HNCO \citep{Noble2015, Fedoseev2015a, Fedoseev2016}. Theoretical calculations have suggested that the link of HNCO with formamide may stem from the two molecules reacting analogously in a physical environment at a certain temperature, but not necessarily implying a chemical connection \citep{Quenard2018a}.

\subsection{Contributions from disc chemistry}
The emission from gas-phase molecules that are observed with ALMA in IRAS~16293-2422~B is assumed to be directly representative of the ices that are being transported into the `disc-like' structure around the protostar. However, this may not necessarily be the case as a result of the precise transport mechanisms of planetesimals and cometesimals into a protoplanetary disc. The exact location of formation of such bodies cannot be observed directly, and neither can their route into a disc. Only theoretical studies can probe these physical processes and suggest that the enhanced dust temperatures and UV fluxes do chemically alter the volatiles between the prestellar and protoplanetary disc phases during infall \citep{Visser2009, Visser2011, Drozdovskaya2014, Drozdovskaya2016, Hincelin2016, Yoneda2016}. However, what exactly transpires at the disc-envelope boundary still remains unclear, for example. On the other hand, once inside the disc, icy volatiles that are locked up in sufficiently large cometesimals and that remain in the outer parts of the protoplanetary disc for the rest of the time, would no longer be affected by disc chemical processes. Hence, implying that the bulk composition of cometesimals could still be pristine disc-composing materials.

Direct observations of ices in protostellar systems and protoplanetary discs would be more directly comparable to cometary volatiles. Unfortunately, solid state observations have only been possible thanks to unique configurations in a handful of somewhat older (Class II) discs with only H$_{2}$O, CO, OCN$^{-}$, OCS, and tentatively HDO being detected so far \citep{Pontoppidan2005, Honda2009, TeradaTokunaga2012, Terada2012, Aikawa2012, McClure2015}. The James Webb Space Telescope (JWST) is expected to make much greater progress on this topic (for example within the framework of the Mid-Infrared Instrument (MIRI) European Consortium (EC) ``Protostars Survey'' Guaranteed Time Observations (GTO) program, PI: Ewine F. van Dishoeck, and the ``IceAge: Chemical Evolution of Ices during Star Formation'' Director’s Discretionary Early Release Science (DD-ERS) program, \citealt{McClure2018}). Such comparisons will be the subject of future work; however, they will always be limited to the most abundant icy volatiles due to the need for large quantities of individual molecules to generate detectable absorption features. The full chemical inventory of such diverse sets of molecules is only possible in the gas phase with facilities such as ALMA.

%%%%%%%%%%%%%%%%%%%%%%%%%%%%%%%%%%%%%%%%%%%%%%%%%%%%%%%%%%%%%%%%%%%%%%%%%%%%%%%
\section{Conclusions}
\label{conclusions}

In the quest to identify the ingredients that are needed to form Solar-like systems, a comparative study has been carried out between IRAS~16293-2422~B and 67P/C--G. IRAS~16293-2422 is an embedded low-mass binary protostellar system that is thought to be analogous to the youngest stages of formation of our Solar System. Source B is favorably positioned in the sky for a complete and quantitative chemical inventory with observations carried out by ALMA with the PILS survey on protoplanetary disc-scales \citep{Jorgensen2016}. 67P/C--G is a Jupiter-family comet that has been monitored continuously for more than 2 years by the instruments of the \textit{Rosetta} mission allowing an unprecedented characterization of its composition and the first-time derivation of bulk cometary molecular abundances \citep{Rubin2019a}. In this paper, the most complete molecular inventory to date of both targets has been compared in terms of relative abundances. The main conclusions are as follows.

\begin{enumerate}
	\item Abundances of CHO-, N- and S-bearing molecules display correlations between the protostellar IRAS~16293-2422~B and the cometary 67P/C--G volatiles relative to CH$_{3}$OH, CH$_{3}$CN, and CH$_{3}$SH, respectively, with some scatter. Tentative correlations between P- and Cl-bearing molecules relative to CH$_{3}$OH are inferred. This suggests preservation of prestellar and protostellar volatiles into cometary bodies upon some degree of chemical alteration.
	\item Cometary relative abundances (as measured for 67P/C--G) tend to be higher than protostellar quantities (as observed in IRAS~16293-2422~B) for CHO- and N-bearing species, which may indicate either that volatile molecules are destroyed near the protostar before entry into the protoplanetary disc or that more have been produced by the time of incorporation into the comet. It cannot be excluded that this may stem from variations of solely the reference molecules (CH$_{3}$OH and CH$_{3}$CN) between comets and protostellar regions.
	\item Links between Hale--Bopp's and ISM volatiles have been confirmed for the case of 67P/C--G for CHO- and N-bearing molecules on protoplanetary disc-scales. For S-bearing species these may have been missed previously for Hale--Bopp due to the use of an unrepresentative reference molecule, the importance of distributed sources for S-bearing volatiles, the snap-shot nature of cometary ground-based observations or the low spatial resolution ISM data points that encompass many structures of star-forming regions simultaneously.
	\item The volatile composition of cometesimals and planetesimals is partially inherited from the pre- and protostellar phases of evolution.
\end{enumerate}

A more direct comparison with bulk cometary volatiles could be achieved by probing protoplanetary disc ices with data from future mission such as the JWST; however, this would always be limited to only the most-abundant icy species, as minor constituents would not generate observable absorption features. The legacy of the detailed in situ study of a comet as was achieved with the \textit{Rosetta} mission should be extended in the future through analogous missions to comets of different dynamic origins and other small bodies of our Solar System.

%%%%%%%%%%%%%%%%%%%%%%%%%%%%%%%%%%%%%%%%%%%%%%%%%%%%%%%%%%%%%%%%%%%%%%%%%%%%%%%
\section{Acknowledgements}
\label{acknowledgements}

This work is supported by the Swiss National Science Foundation (SNSF) Ambizione grant 180079, the Center for Space and Habitability (CSH) Fellowship and the IAU Gruber Foundation Fellowship. MR acknowledges the support of the state of Bern and the SNSF (200020\_182418). JKJ is supported by the European
Research Council (ERC) under the European Union’s Horizon 2020 research and innovation programme through ERC Consolidator Grant “S4F” (grant agreement
No. 646908). Research at Centre for Star and Planet Formation is funded by the Danish National Research Foundation.

The authors would like to acknowledge the contributions to this work of the entire PILS and ROSINA teams, as well as input of Holger~S.~P.~M\"{u}ller with regards to the spectroscopy of glycine, and useful discussions with Nadia Murillo and Matthijs van der Wiel about binary protostellar sources.

This paper makes use of the following ALMA data: ADS/JAO.ALMA\#2013.1.00278.S, ADS/JAO.ALMA\#2012.1.00712.S, ADS/JAO.ALMA\#2016.1.01150.S., ADS/JAO.ALMA\#2011.0.00007.SV, ADS/JAO.ALMA\#2013.1.00061.S, ADS/JAO.ALMA\#2017.1.00518.S, and ADS/JAO.ALMA\#2015.1.01193.S. ALMA is a partnership of ESO (representing its member states), NSF (USA) and NINS (Japan), together with NRC (Canada), MOST and ASIAA (Taiwan), and KASI (Republic of Korea), in cooperation with the Republic of Chile. The Joint ALMA Observatory is operated by ESO, AUI/NRAO and NAOJ.

%%%%%%%%%%%%%%%%%%%%%%%%%%%%%%%%%%%%%%%%%%%%%%%%%%%%%%%%%%%%%%%%%%%%%%%%%%%%%%%
\clearpage
\bibliographystyle{mn2e}
\bibliography{mybib} % mybib.bib file

%%%%%%%%%%%%%%%%%%%%%%%%%%%%%%%%%%%%%%%%%%%%%%%%%%%%%%%%%%%%%%%%%%%%%%%%%%%%%%%
\clearpage
\newpage
\appendix

%%%%%%%%%%%%%%%%%%%%%%%%%%%%%%%%%%%%%%%%%%%%%%%%%%%%%%%%%%%%%%%%%%%%%%%%%%%%%%%
\section{Newly determined column densities in IRAS~16293-2422~B}
\label{columndens_IRAS16293B}

This study required a full chemical inventory of IRAS~16293-2422~B, including some molecules for which column densities (or at least upper limits on them) have not been previously published, and hence, had to be newly derived. These are presented in this appendix. The synthetic spectra have been generated with custom IDL routines under the assumption of local thermal equilibrium (LTE), which has been justified for the case of methanol in section~$5.1$ of \citet{Jorgensen2016}. The spectroscopic data stem from the Cologne Database of Molecular Spectroscopy (CDMS; \citealt{Muller2001, Muller2005, Endres2016}).

\subsection{CO}
For a highly abundant species, such as CO, line optical thickness is a severe problem for determining the column densities, even of the isotopologues. The PILS Band 7 range covers the $J=3-2$ transitions of the four main isotopologues of CO ($^{12}$CO, $^{13}$CO, C$^{18}$O and C$^{17}$O); while the transitions of the less abundant doubly substituted variants and the vibrationally excited states fall at lower frequencies outside of this range. The two rarer variants of these four, C$^{18}$O, and C$^{17}$O, both show inverse P Cygni profiles towards the one-beam offset position, which is an indication of significant optical depth. Consequently, only a ball-park number for the CO column density can be derived: for C$^{17}$O, the blue emission part of the line profiles suggests temperatures in the $100-150$~K range with a column density of $5\times10^{16}$~cm$^{-2}$. A higher column density can also fit the C$^{17}$O transition, if a lower excitation temperature is adopted. However, that would be at odds with the C$^{18}$O $J=3-2$ transition that is then severely under-produced.

The derived C$^{17}$O column density would imply an overall CO column density of $\sim1\times10^{20}$~cm$^{-2}$, i.e., an order of magnitude above that of CH$_3$OH. Adopting the lower limit for the H$_{2}$ column density based on the dust continuum emission \citep{Jorgensen2016}, the upper limit to the CO abundance relative to H$_{2}$, is $2\times10^{-5}$. These relative CH$_{3}$OH/CO and CO/H$_{2}$ abundances are not unreasonable, e.g., in comparison to typical ice measurements in star-forming regions. However, due to the above uncertainties, the derived value is at best only accurate to within a factor of a few.

\subsection{HCN and HNC}
Like for CO, the lines of the HCN and HNC are strongly affected by optical depth issues. The PILS Band 7 data cover the HC$^{15}$N $J=4-3$ transition at $344.2$~GHz. If an excitation temperature of $120$~K is assumed, this transition is fit by a column density of $2.5\times10^{14}$~cm$^{-2}$ with the emission being marginally optically thick. This corresponds to a column density of $5\times16$~cm$^{-2}$ for the main isotopologue. The PILS Band 7 data also cover the $J=4-3$ transition of DC$^{15}$N and a tentative assignment can be made at an implied column density of $7.5\times10^{12}$~cm$^{-2}$. This would correspond to a D/H ratio of about $3$ per cent, which is in-line with other measurements of D/H ratios from PILS (e.g., \citealt{Drozdovskaya2018, Jorgensen2018}). Conversely, if the HCN column density had been grossly underestimated, the D/H ratio would need to be much lower for this species in particular. Finally, the PILS Band 7 data also cover two lines of vibrationally excited HCN. If correctly assigned, these transitions would imply a column density of 3$\times10^{17}$~cm$^{-2}$, assuming an excitation temperature of $120$~K, while an assumed temperature of $200$~K would lower this to 2$\times10^{16}$~cm$^{-2}$. Thus, for HCN, the vibrationally excited lines correspond to the same regime as the isotopologues, but do not provide any stronger constraints.

For HNC, no emission is seen from either isotopologue or vibrationally excited states. A possible explanation is that HNC has a much more compact distribution than that of HCN, causing its lines to be completely optically thick and fully quenched. It is thought that its column density cannot be higher than that of HCN.

Additionally, the detected $J=1-0$ lines of HNC and HCN lines in ALMA Band~$3$ were examined (project-id 2017.1.00518.S, PI: Ewine F. van Dishoeck, and project-id 2015.1.01193.S, PI: V{\'i}ctor M. Rivilla), which indicate that the emission from these low excitation lines stems primarily from the circumbinary envelope scales. These data were obtained at a lower spatial resolution than the PILS observations ($\sim1\arcsec$ versus $0\farcs{5}$), making it hard to conclusively derive the emitting regions. The analysis of HNC and HCN lines in ALMA Band~$7$ (PILS) suggests a much smaller column density for HNC than that of HCN, as no emission is seen at frequencies corresponding to lines of vibrationally excited HNC and its isotopologues, while lines of vibrationally excited HCN and its isotopologues are detected. Consequently, an upper limit on $N(\text{HNC})$ equal to $N(\text{HCN})$ is assumed in this work.

\subsection{PN}
\begin{figure}
 \centering
  \includegraphics[width=0.45\textwidth,height=0.8\textheight,keepaspectratio]{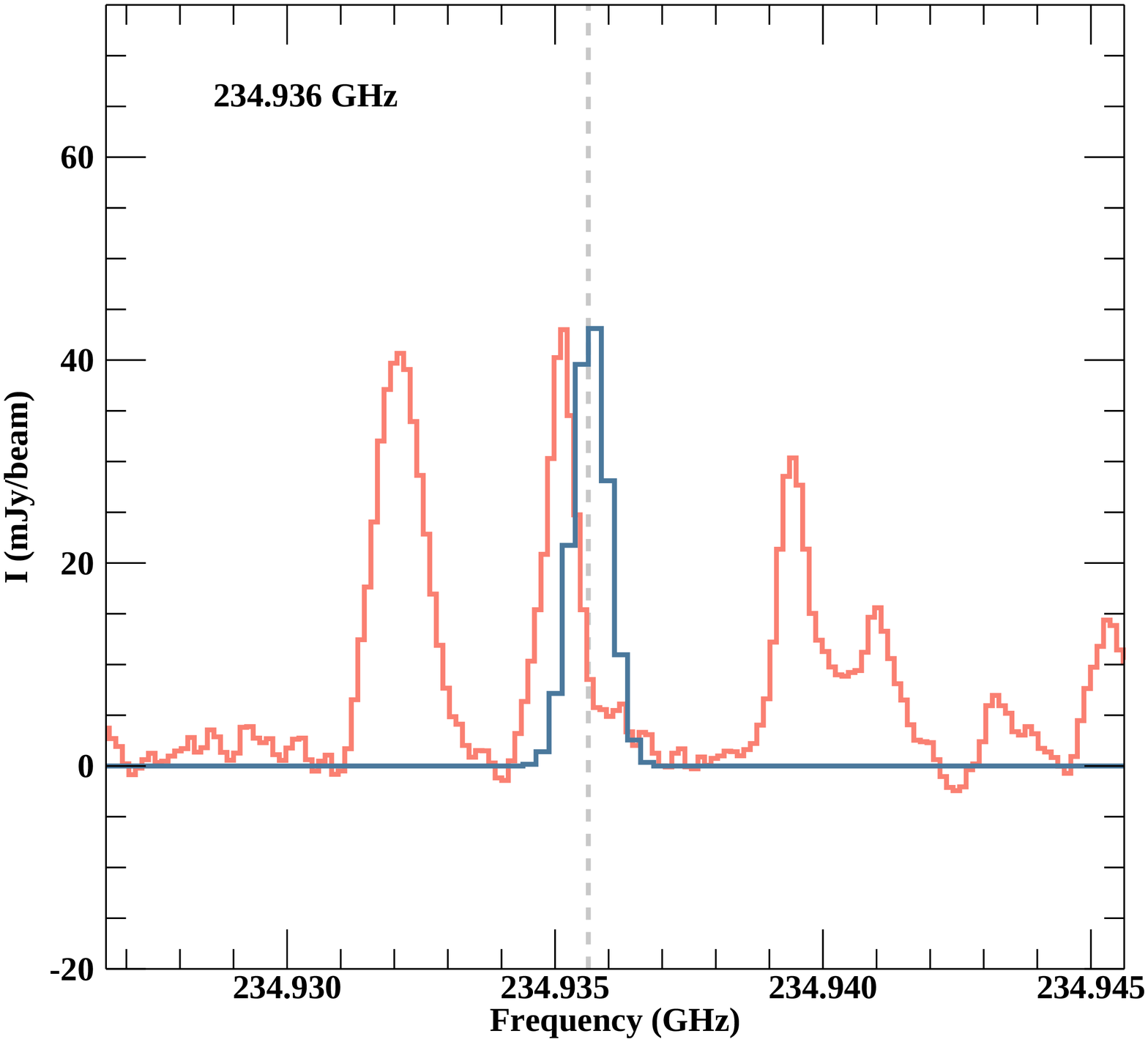}
 \caption{The detected $J=5-4$ line of PN in the $v=0$ state (CDMS entry 045511) with $E_{\text{up}}=34$~K and $A_{ij}=5.2\times10^{-4}$~s$^{-1}$. The observed ALMA Band 6 spectrum convolved with a uniform circular restoring beam of $0\farcs{5}$ at the one-beam ($\sim70$~au) offset from source B of IRAS 16293–2422 in the SW direction is in salmon. The turquoise line is the LTE fit for the transition, assuming a source size that is equal to the beam size of $0\farcs{5}$, full width half-maximum (FWHM) of $1$~km~s$^{-1}$, $T_{\text{ex}}=125$~K and $N=2.1\times10^{13}$~cm$^{-2}$. The local standard of rest (LSR) velocity, $v_{\text{LSR}}$, is assumed to be $2.7$~km~s$^{-1}$ at this position.}
 \label{fgr:PN6}
\end{figure}

\begin{figure}
 \centering
  \includegraphics[width=0.45\textwidth,height=0.8\textheight,keepaspectratio]{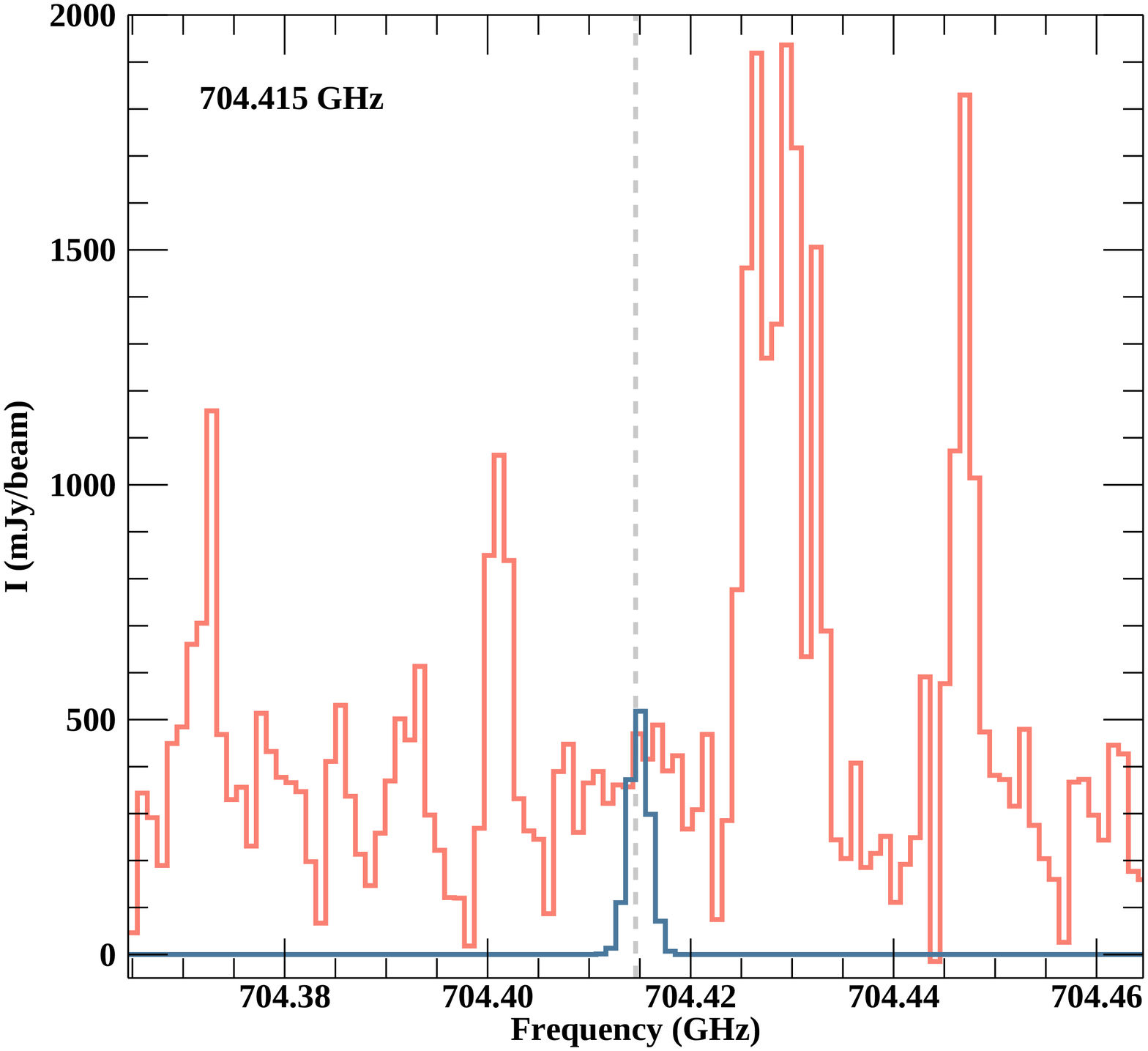}
 \caption{The covered $J=15-14$ line of PN in the $v=0$ state (CDMS entry 045511) with  $E_{\text{up}}=271$~K and $A_{ij}=1.5\times10^{-2}$~s$^{-1}$. The observed ALMA Band 9 spectrum convolved with a uniform circular restoring beam of $0\farcs{5}$ at the one-beam ($\sim70$~au) offset from source B of IRAS 16293–2422 in the SW direction is in salmon. The turquoise line is the LTE fit for the transition, assuming a source size that is equal to the beam size of $0\farcs{5}$, FWHM of $1$~km~s$^{-1}$, $T_{\text{ex}}=125$~K and $N=2.1\times10^{13}$~cm$^{-2}$. $v_{\text{LSR}}=2.7$~km~s$^{-1}$ is assumed at this position.}
 \label{fgr:PN9}
\end{figure}

IRAS~16293-2422 was observed in ALMA Band 6 under project-id 2016.1.01150.S (PI: Vianney Taquet) in Cycle 4. The dataset appears to contain a detection of the $J=5-4$ line of PN in the $v=0$ state (Figure~\ref{fgr:PN6}) with $E_{\text{up}}=34$~K and $A_{ij}=5.2\times10^{-4}$~s$^{-1}$. This line was also covered in project-id: 2016.1.00457, PI: Yoko Oya; however, the dataset has not been compared with that of project-id 2016.1.01150.S, PI: Vianney Taquet. The binary system was also observed in ALMA Band 9 under project-id 2011.0.00007.SV during the Science Verification phase. The dataset covers the $J=15-14$ line of PN in the $v=0$ state (Figure~\ref{fgr:PN9}) with $E_{\text{up}}=271$~K and $A_{ij}=1.5\times10^{-2}$~s$^{-1}$. However, this dataset suffers from a poor baseline and a high noise level, making it hard to firmly constrain $T_{\text{ex}}$ and $N$. It can only be used to verify that the column density derived based on the Band 6 line does not contradict the emission observed in Band 9. Using the spectra upon convolution with a uniform circular restoring beam of $0\farcs{5}$, yields $N(\text{PN})=2.1\times10^{13}$~cm$^{-2}$ at the one-beam ($\sim70$~au) offset from source B of IRAS 16293–2422 in the SW direction upon assuming $T_{\text{ex}}=125$~K. The observed $J=5-4$ line of PN should not be blended with any other species, based on checks with currently available spectroscopic catalogues. The secondary minor emission bump to the right of the line may stem from two weak acetone lines. The observed frequency shift of about two spectral channels may be due to a velocity shift with respect to the assumed source velocity, which may point towards PN emission tracing a different, adjacent component of the system. Upcoming new dedicated ALMA observations will secure the detection and firmly constrain the column density, excitation temperature and associated velocity (project-id: 2018.1.01496, PI: V{\'i}ctor M. Rivilla).

The PILS data in Bands 7 and 3 cover two lines of P$^{15}$N ($358.686$ and $89.685$~GHz, with the Band 7 line being also covered in the continuum window of project-id: 2017.1.00568, PI: Lars Kristensen). Two more lines are covered in Bands 7 and 8 of project-id: 2013.1.00061, PI: Audrey Coutens. No emission is seen from P$^{15}$N in any of these four lines, although synthetic spectra predict these to be very weak ($<2$~mJy/beam, assuming LTE, a source size that is equal to the beam size of $0\farcs{5}$, FWHM of $1$~km~s$^{-1}$, $T_{\text{ex}}=125$~K and $N=1.5\times10^{11}$~cm$^{-2}$).

\clearpage
\newpage

\subsection{PO}
\begin{figure*}
 \centering
 \begin{subfigure}[b]{0.45\textwidth}
  \includegraphics[width=\textwidth,height=0.8\textheight,keepaspectratio]{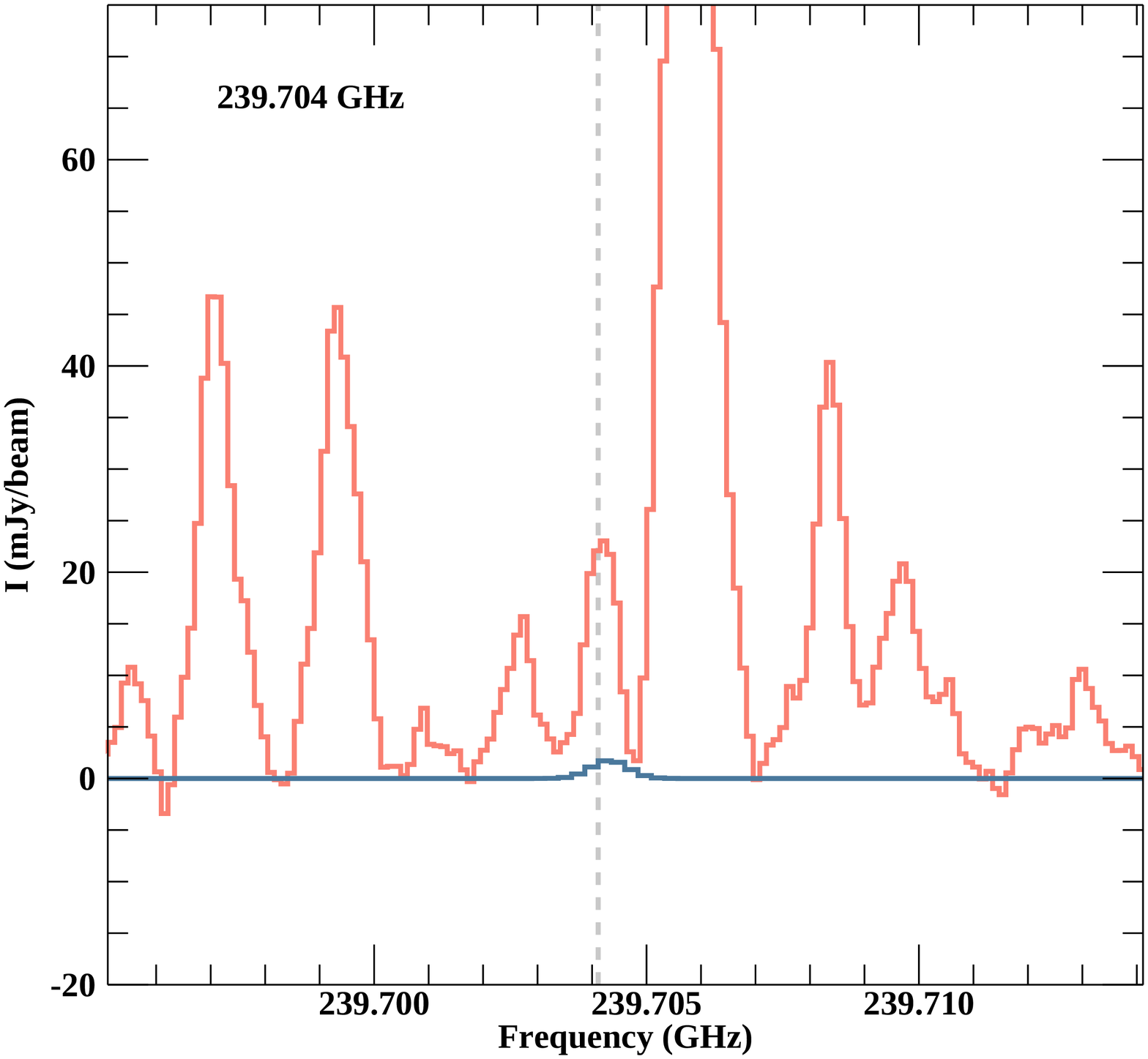}
 \end{subfigure}
 \begin{subfigure}[b]{0.45\textwidth}
  \includegraphics[width=\textwidth,height=0.8\textheight,keepaspectratio]{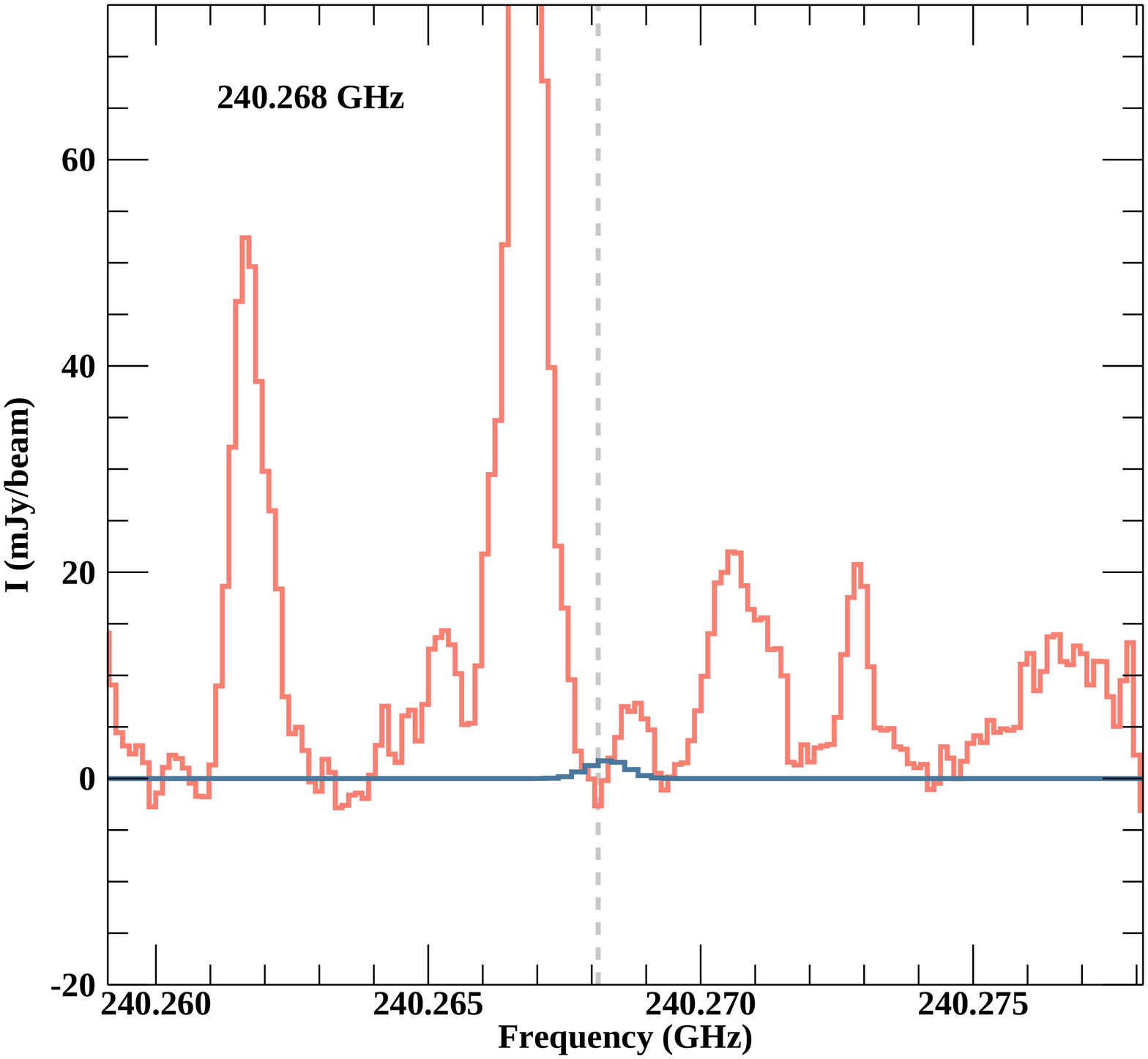}
 \end{subfigure}
 \caption{The covered lines of PO (CDMS entry 047507) with $E_{\text{up}}=37$~K and $A_{ij}=4.7\times10^{-6}$~s$^{-1}$. The observed ALMA Band 6 spectrum convolved with a uniform circular restoring beam of $0\farcs{5}$ at the one-beam ($\sim70$~au) offset from source B of IRAS 16293–2422 in the SW direction is in salmon. The turquoise line is the LTE fit for the transition, assuming a source size that is equal to the beam size of $0\farcs{5}$, FWHM of $1$~km~s$^{-1}$, $T_{\text{ex}}=125$~K and $N=4.4\times10^{14}$~cm$^{-2}$. $v_{\text{LSR}}=2.7$~km~s$^{-1}$ is assumed at this position.}
 \label{fgr:POB6}
\end{figure*}

\begin{figure*}
 \centering
  \includegraphics[width=\textwidth,height=0.8\textheight,keepaspectratio]{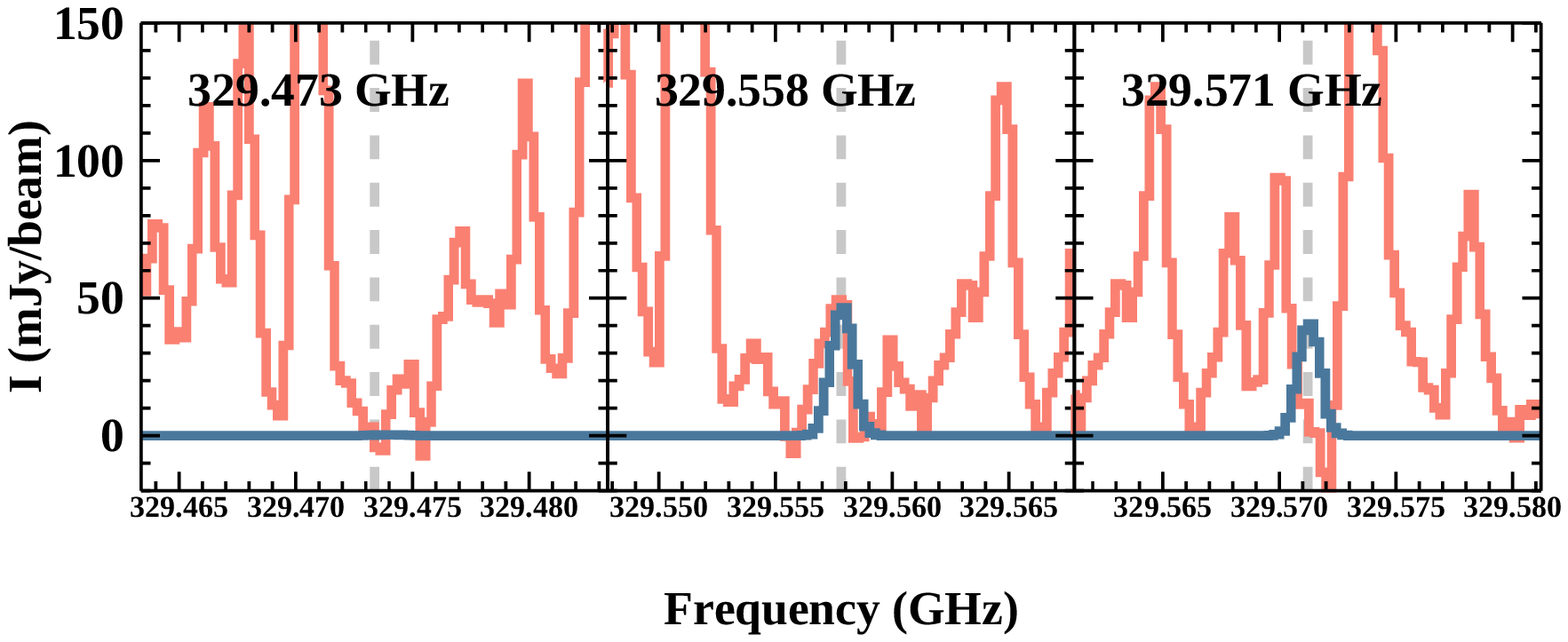}
 \caption{The covered lines of PO (CDMS entry 047507) with $E_{\text{up}}=387$~K and $A_{ij}=6.3\times10^{-6}$ and $6.6\times10^{-4}$~s$^{-1}$ for the left and middle jointly with right panels, respectively. The observed ALMA Band 7 spectrum convolved with a uniform circular restoring beam of $0\farcs{5}$ at the one-beam ($\sim70$~au) offset from source B of IRAS 16293–2422 in the SW direction is in salmon. The turquoise line is the LTE fit for the transition, assuming a source size that is equal to the beam size of $0\farcs{5}$, FWHM of $1$~km~s$^{-1}$, $T_{\text{ex}}=125$~K and $N=4.4\times10^{14}$~cm$^{-2}$. $v_{\text{LSR}}=2.7$~km~s$^{-1}$ is assumed at this position. The middle panel shows a tentative detection, while the left panel contains a non-detected line due to intrinsic line weakness and the right panel contains a non-detection due to interference with the absorption feature caused by a strong quadrupole line of dimethyl ether to the right.}
 \label{fgr:POB7}
\end{figure*}

Two weak ($A_{ij}=4.7\times10^{-6}$~s$^{-1}$) lines of PO are covered by the PILS data in Band 6 with $E_{\text{up}}=37$~K ($239.704$ and $240.268$~GHz, with the latter line being also covered in project-id: 2016.1.00457, PI: Yoko Oya). Synthetic spectra predict these two lines to be of equal strength; however, in the observations one is completely absent (Figure~\ref{fgr:POB6}). This makes it rather unlikely that the other observed line stems from PO. The observed emission at this frequency likely stems from vinyl cyanide and CH$_{2}$DOH. Furthermore, synthetic spectra predict these to be very weak ($<2$~mJy/beam), assuming LTE, a source size that is equal to the beam size of $0\farcs{5}$, FWHM of $1$~km~s$^{-1}$, $T_{\text{ex}}=125$~K and $N=4.4\times10^{14}$~cm$^{-2}$. More lines of PO are covered by the PILS data in Band 7: six in total; however, giving only three spectrally resolved lines due to the proximity in frequency. Synthetic spectra predict the observed emission to be $\sim0.4$, $48$, and $42$~mJy/beam, respectively (Fig.~\ref{fgr:POB7}). The data show a tentative detection at $329.557$~GHz, which should not be blended with any other species, based on checks with currently available spectroscopic catalogues. Synthetic spectra predict a detectable line at $329.571$~GHz as well; however, there is a very strong quadruple dimethyl ether line to the right, which is causing an absorption feature in the observed spectrum. This absorption likely prevents the PO emission from being seen at this frequency. Consequently, only an upper limit is derived for the PO column density of $4.4\times10^{14}$~cm$^{-2}$ at the one-beam ($\sim70$~au) offset from source B of IRAS 16293–2422 in the SW direction.

Five more very weak PO lines have been covered by ALMA observations of IRAS~16293-2422; however, synthetic spectra predict all these to be below $\sim0.1$~mJy/beam in intensity and consequently, unobservable. Specifically, there are two lines at $\sim241.574$~GHz in Band 6 (project-id: 2013.1.00061, PI: Audrey Coutens); one line at $\sim349.792$~GHz in Band 7 (covered by the PILS survey and by project-id: 2013.1.00018, PI: Markus Schmalzl)\footnote{Note that this line is only available through the Jet Propulsion Laboratory (JPL) catalogue (\citealt{Pickett1998}; entry 47006) due to its intrinsic weakness and the intensity cut-off used by the CDMS catalogue.}; and a double line at $\sim109.830$~GHz (project-id: 2017.1.01247, PI: Giovanni Dipierro). Finally, ten lines of PO lie in the continuum window of project-id: 2016.1.01468, PI: Victor Magalh$\tilde{\text{a}}$es, but the strongest of which is predicted to be only $\sim4$~MJy/beam. As for PN, upcoming new dedicated ALMA observations will secure the detection and firmly constrain the column density and excitation temperature (project-id: 2018.1.01496, PI: V{\'i}ctor M. Rivilla).

\clearpage
\newpage

\subsection{Glycine}
Spectroscopic characterization of glycine is associated with many large uncertainties. Uncertainties in the ALMA Band~$7$ frequency range are thought to be at least $\sim0.15-0.25$~MHz for the strongest lines. The molecule has many low-lying modes, which increase the vibrational correction factor, especially at $T_{\text{ex}}\sim300$~K. Gas-phase measurements are currently unavailable, only theoretical calculations can be found in the literature. Crude estimates for the vibrational correction factors for Conformer I are $2.05$ and $10.65$ at $125$~K and $300$~K, respectively. For example, the consideration of an additional low-lying conformer at $1.4\pm0.4$~kcal/mol would introduce an additional correction factor of $1.1-1.5$ at $T_{\text{ex}}\sim300$~K (H. M{\"u}ller priv. comm.).

\begin{figure*}
 \centering
  \includegraphics[width=0.8\textwidth,height=0.8\textheight,keepaspectratio]{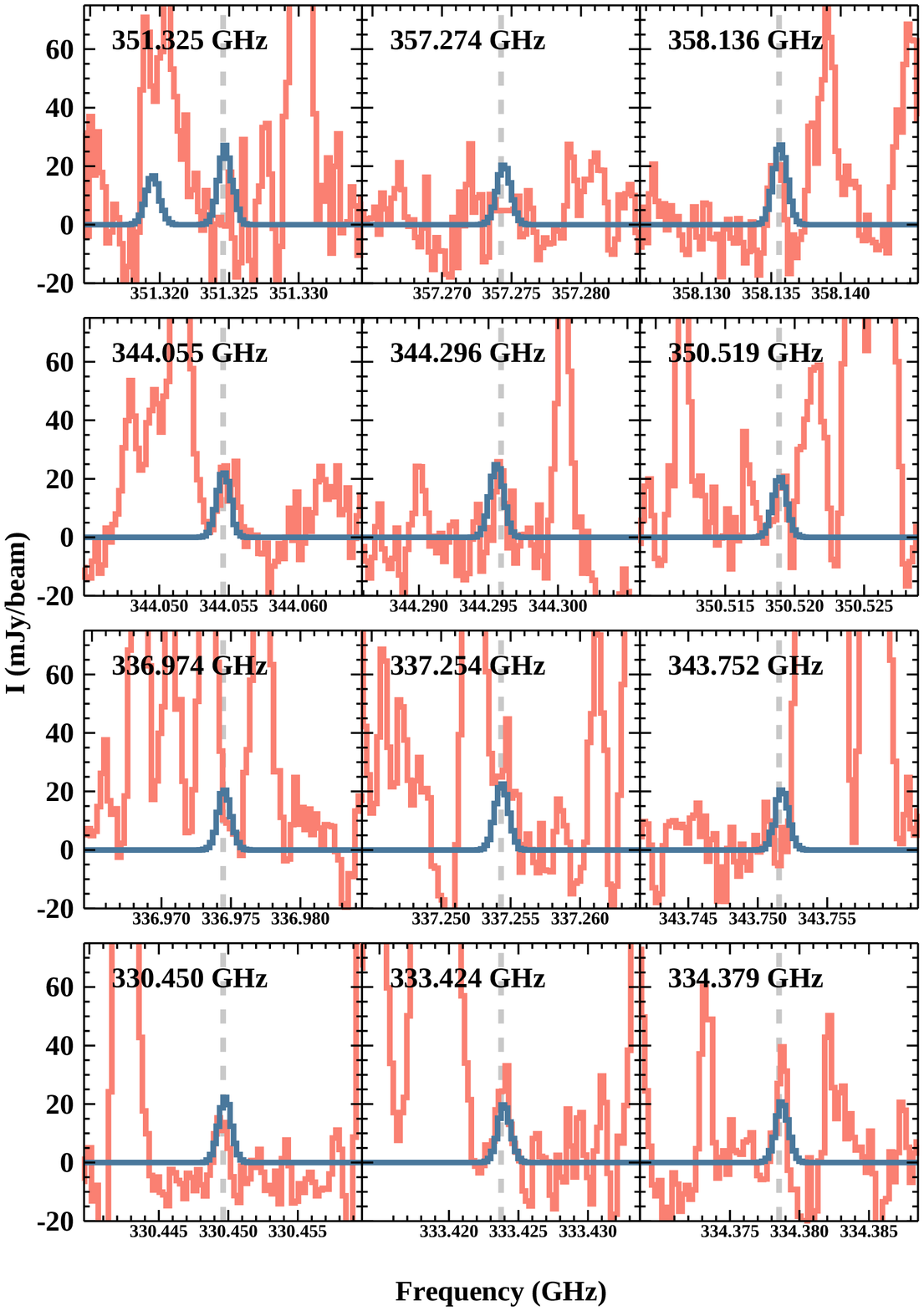}
 \caption{The synthetic spectra overlaid onto observed spectra around $12$ of the strongest predicted lines of glycine Conformer I (CDMS entry 075511) in the PILS frequency range. The observed PILS ALMA Band 7 spectrum convolved with a uniform circular restoring beam of $0\farcs{5}$ at the one-beam ($\sim70$~au) offset from source B of IRAS 16293–2422 in the SW direction is in salmon. The turquoise lines are the LTE fits for the covered transitions, assuming a source size that is equal to the beam size of $0\farcs{5}$, FWHM of $1$~km~s$^{-1}$, $T_{\text{ex}}=125$~K and $N\leq3.0\times10^{15}$~cm$^{-2}$ (upon accounting for the vibrational correction factor and the correction factor for the coupling of line emission with the emission from dense dust at $T_{\text{bg}}$). $v_{\text{LSR}}=2.7$~km~s$^{-1}$ is assumed at this position.}
 \label{fgr:GlyI125K}
\end{figure*}

\begin{figure*}
 \centering
  \includegraphics[width=0.8\textwidth,height=0.8\textheight,keepaspectratio]{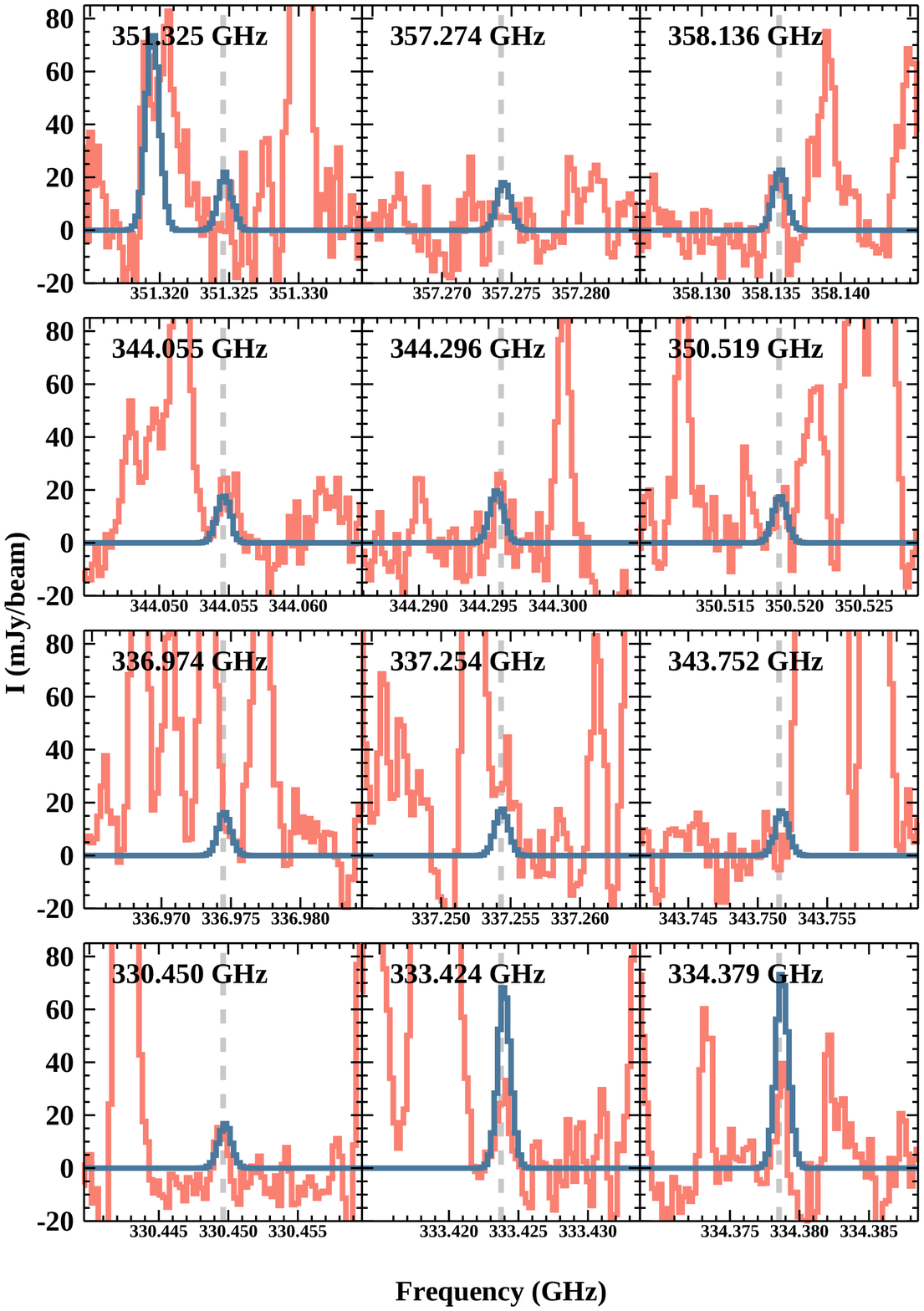}
 \caption{The synthetic spectra overlaid onto observed spectra around $12$ of the strongest predicted lines of glycine Conformer I (CDMS entry 075511) in the PILS frequency range. The observed PILS ALMA Band 7 spectrum convolved with a uniform circular restoring beam of $0\farcs{5}$ at the one-beam ($\sim70$~au) offset from source B of IRAS 16293–2422 in the SW direction is in salmon. The turquoise lines are the LTE fits for the covered transitions, assuming a source size that is equal to the beam size of $0\farcs{5}$, FWHM of $1$~km~s$^{-1}$, $T_{\text{ex}}=300$~K and $N\leq1.3\times10^{14}$~cm$^{-2}$ (upon accounting for the vibrational correction factor and the correction factor for the coupling of line emission with the emission from dense dust at $T_{\text{bg}}$). $v_{\text{LSR}}=2.7$~km~s$^{-1}$ is assumed at this position.}
 \label{fgr:GlyI300K}
\end{figure*}

\begin{figure*}
 \centering
  \includegraphics[width=0.8\textwidth,height=0.8\textheight,keepaspectratio]{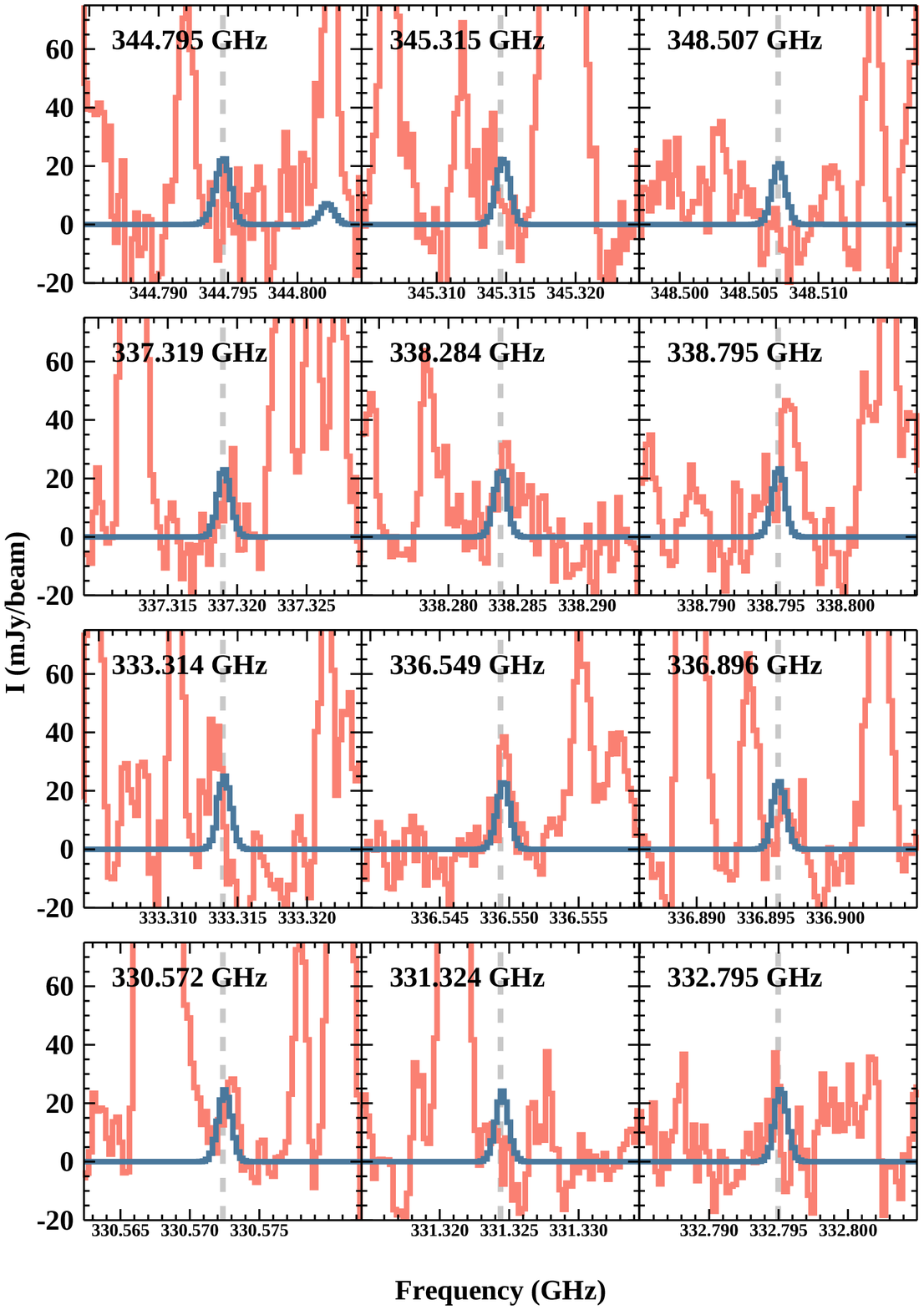}
 \caption{The synthetic spectra overlaid onto observed spectra around $12$ of the strongest predicted lines of glycine Conformer II (CDMS entry 075512) in the PILS frequency range. The observed PILS ALMA Band 7 spectrum convolved with a uniform circular restoring beam of $0\farcs{5}$ at the one-beam ($\sim70$~au) offset from source B of IRAS 16293–2422 in the SW direction is in salmon. The turquoise lines are the LTE fits for the covered transitions, assuming a source size that is equal to the beam size of $0\farcs{5}$, FWHM of $1$~km~s$^{-1}$, $T_{\text{ex}}=125$~K and $N\leq8.9\times10^{14}$~cm$^{-2}$ (upon accounting for the vibrational correction factor and the correction factor for the coupling of line emission with the emission from dense dust at $T_{\text{bg}}$). $v_{\text{LSR}}=2.7$~km~s$^{-1}$ is assumed at this position.}
 \label{fgr:GlyII125K}
\end{figure*}

\begin{figure*}
 \centering
  \includegraphics[width=0.8\textwidth,height=0.8\textheight,keepaspectratio]{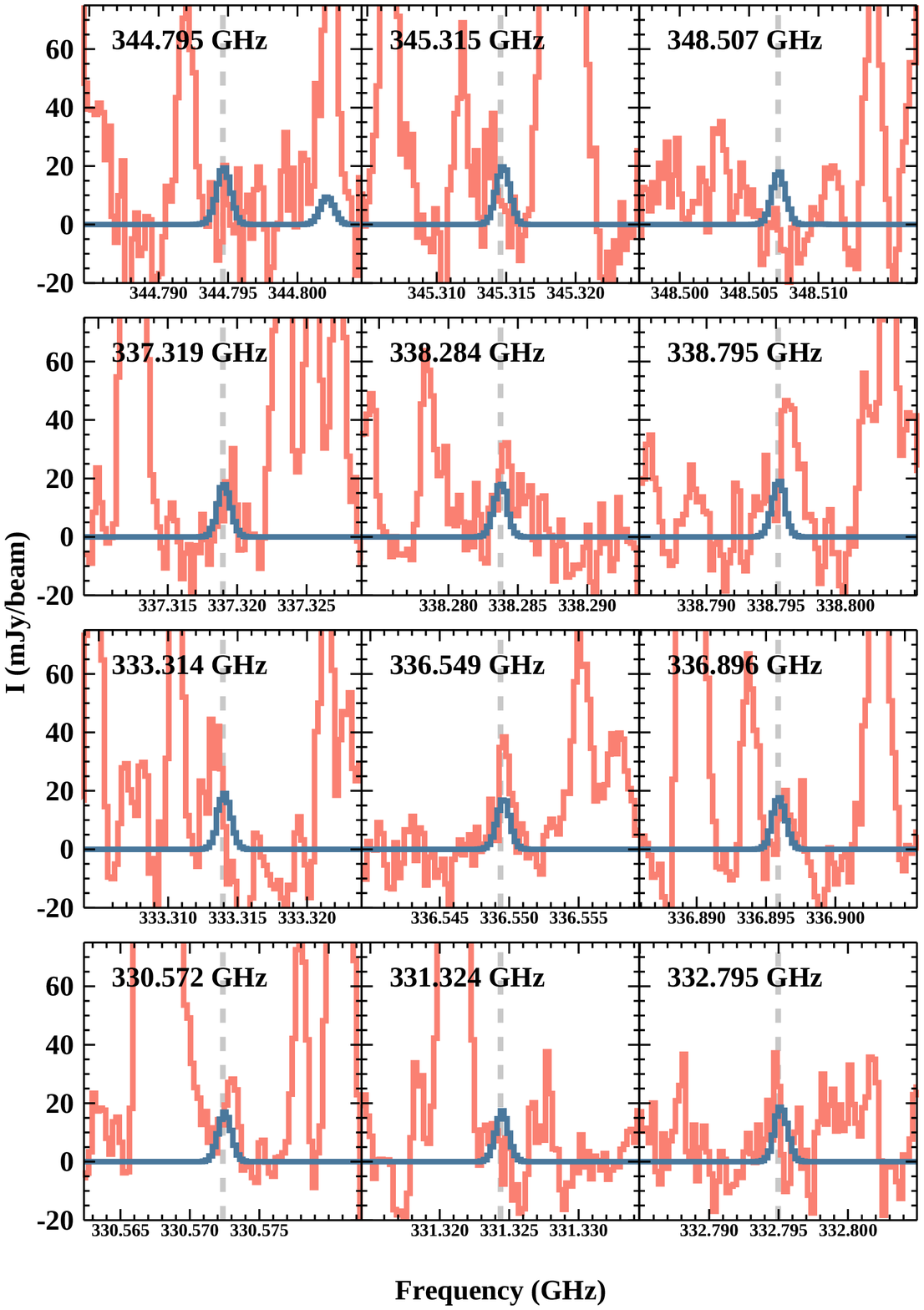}
 \caption{The synthetic spectra overlaid onto observed spectra around $12$ of the strongest predicted lines of glycine Conformer II (CDMS entry 075512) in the PILS frequency range. The observed PILS ALMA Band 7 spectrum convolved with a uniform circular restoring beam of $0\farcs{5}$ at the one-beam ($\sim70$~au) offset from source B of IRAS 16293–2422 in the SW direction is in salmon. The turquoise lines are the LTE fits for the covered transitions, assuming a source size that is equal to the beam size of $0\farcs{5}$, FWHM of $1$~km~s$^{-1}$, $T_{\text{ex}}=300$~K and $N\leq2.7\times10^{13}$~cm$^{-2}$ (upon accounting for the vibrational correction factor and the correction factor for the coupling of line emission with the emission from dense dust at $T_{\text{bg}}$). $v_{\text{LSR}}=2.7$~km~s$^{-1}$ is assumed at this position.}
 \label{fgr:GlyII300K}
\end{figure*}

The PILS Band~$7$ frequency range covers $949$ lines of Conformer I and $1085$ lines of Conformer II when making use of the CDMS entries 075511 and 075512, respectively. These have been used to estimate upper limits on the column density of glycine for excitation temperatures of $125$ and $300$~K. The obtained values are tabulated in Table~\ref{tbl:Ngly} upon applying the vibrational correction factors at these two temperatures as derived for Conformer I to both conformers. The synthetic spectra overlaid onto observed spectra around $12$ of the strongest predicted lines of each conformer in the PILS frequency range are shown in Figs.~\ref{fgr:GlyI125K}-\ref{fgr:GlyII300K}. The column density of glycine used in Table~\ref{tbl:abunvalues_B} is the sum of the two conformers at $T_{\text{ex}}=300$~K, as glycine is expected to be co-spatial with and under comparable excitation conditions as other large complex organic molecules such as glycolaldehyde, ethylene glycol, and formamide.

\ctable[
 caption = {Estimated upper limits on glycine column densities towards IRAS~16293-2422~B\tmark.},
 label = {tbl:Ngly}
 ]{@{\extracolsep{\fill}}lll}{
 \tnote{The estimates are obtained under the assumptions of LTE, a source size that is equal to the beam size of $0\farcs{5}$, FWHM of $1$~km~s$^{-1}$, and two different assumed values for $T_{\text{ex}}$. Vibrational correction factors and the correction factor for the coupling of line emission with the emission from dense dust at $T_{\text{bg}}$ have been accounted for, as described in the text.}
 }{
 \hline
 Glycine       & N (cm$^{-2}$) for $T_{\text{ex}}=125$~K & N (cm$^{-2}$) for $T_{\text{ex}}=300$~K\T\B\\
 \hline
 Conformer I   & $3.0\times10^{15}$                      & $8.9\times10^{14}$\T\\
 Conformer II  & $1.3\times10^{14}$                      & $2.7\times10^{13}$\B\\
\hline}

\clearpage
\newpage

\section{Additional merged correlation plots}
\label{merged_corrplots}
This appendix contains three additional correlation plots for the CHO-, N-, S-, P- and Cl-bearing chemical families. One containing the data of Figs.~\ref{fgr:corrplotCHO},~\ref{fgr:corrplotN},~\ref{fgr:corrplotS}, and~\ref{fgr:corrplotother} merged onto the same scales, while preserving the different respective reference species. The other two figures contain the data when using a common reference species for all molecules probed, namely CH$_{3}$OH and H$_{2}$O. Note that the column density of water towards IRAS~16293-2422~B is an estimate based on the value derived for source A (see Table~\ref{tbl:abunvalues_B} for details).

\begin{figure}
 \centering
  \includegraphics[width=0.45\textwidth,height=0.8\textheight,keepaspectratio]{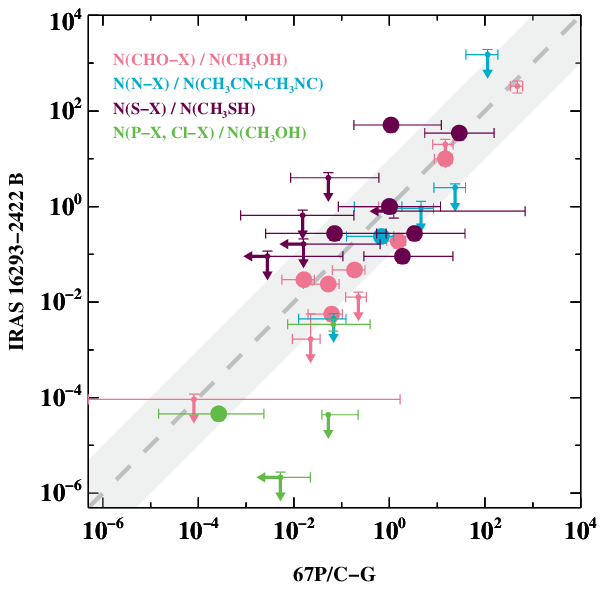}
 \caption{The abundance of CHO-, N-, S-, P- and Cl-bearing molecules relative to methanol, methyl cyanide and methyl mercaptan (depending on the family and as indicated in the legend) detected towards the one-beam offset position from IRAS~16293-2422~B versus that measured in 67P/C--G. Each chemical family is marked with a unique color. The shaded region corresponds to an order of magnitude scatter about the one-to-one linear correlation. Data points that are smaller in size indicate that this pertains to an upper limit or an estimate of sorts either for the protostar, or the comet, or both.}
 \label{fgr:corrplotmerged}
\end{figure}

\begin{figure}
 \centering
  \includegraphics[width=0.45\textwidth,height=0.8\textheight,keepaspectratio]{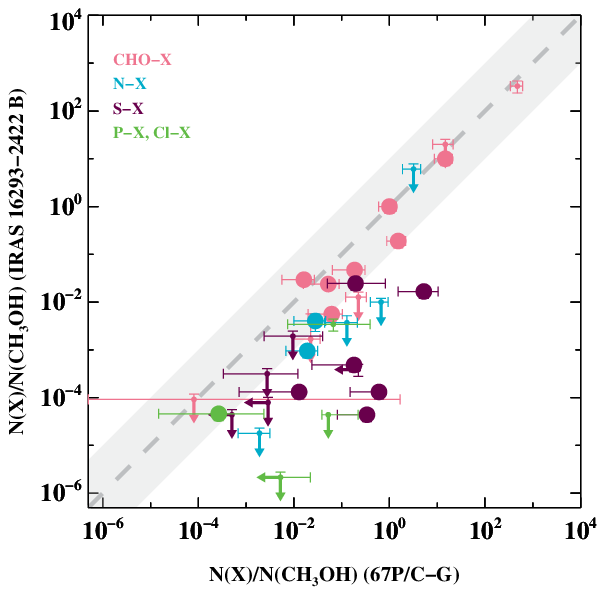}
 \caption{The abundance of CHO-, N-, S-, P- and Cl-bearing molecules relative to methanol detected towards the one-beam offset position from IRAS~16293-2422~B versus that measured in 67P/C--G. Each chemical family is marked with a unique color. The shaded region corresponds to an order of magnitude scatter about the one-to-one linear correlation. Data points that are smaller in size indicate that this pertains to an upper limit or an estimate of sorts either for the protostar, or the comet, or both.}
 \label{fgr:corrplotmergedCH3OH}
\end{figure}

\begin{figure}
 \centering
  \includegraphics[width=0.45\textwidth,height=0.8\textheight,keepaspectratio]{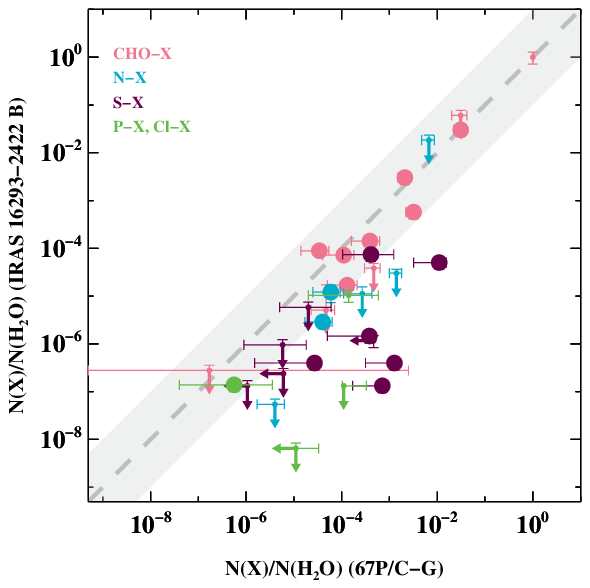}
 \caption{The abundance of CHO-, N-, S-, P- and Cl-bearing molecules relative to water detected towards the one-beam offset position from IRAS~16293-2422~B versus that measured in 67P/C--G. Each chemical family is marked with a unique color. The shaded region corresponds to an order of magnitude scatter about the one-to-one linear correlation. Data points that are smaller in size indicate that this pertains to an upper limit or an estimate of sorts either for the protostar, or the comet, or both.}
 \label{fgr:corrplotmergedH2O}
\end{figure}

\clearpage
\newpage

%%%%%%%%%%%%%%%%%%%%%%%%%%%%%%%%%%%%%%%%%%%%%%%%%%%%%%%%%%%%%%%%%%%%%%%%%%%%%%%
\section{Molecules of IRAS~16293-2422}
\label{molecules_IRAS16293}

In this appendix, the molecules that have been observed towards IRAS~16293-2422 are inventorized. The scales that the respective observations have been carried out on are also considered, and hence, the physical structures that the various species may be tracing are discussed. The cloud scales are $\sim40\arcsec-1\arcmin$, the clump scales are $\sim20-40\arcsec$, the circumbinary envelope scales are $\sim3-20\arcsec$, the individual envelope scales are $\sim1-3\arcsec$ and the disc scales are $\lesssim1\arcsec$. Although, these ranges are somewhat arbitrary as it depends which dust and gas components are being examined. Table~\ref{tbl:abunvalues_B} summarizes the values used for the plots presented in Section~\ref{results}.

\subsection{CHO-bearing molecules}

\subsubsection{Water (H$_{2}$O), $m=18$}
\label{H2O}
The presence of water has long been established for IRAS~16293-2422. It was first observed thanks to masers \citep{WilkingClaussen1987}, which were quickly realized to be associated with source A \citep{Wootten1989, Alves2012} and to also be rapidly variable \citep{Imai1999, Colom2016}. H$_{2}$O and HDO emission has since been detected at radio frequencies on envelope scales \citep{vanDishoeck1995, Stark2004, Caux2011, Lis2016}, and has been claimed to originate from both sources, A and B \citep{Chandler2005, Parise2005, Jorgensen2011}. Para-D$_{2}$O and ortho-D$_{2}$O have been discovered towards the binary \citep{Butner2007, Vastel2010}. H$_{2}$O has been detected in the infrared with ISO/LWS and the Odin satellite \citep{Ceccarelli1998a, Hjalmarson2003}. H$_{2}$O, HDO and ortho-D$_{2}$O lines were also detected on envelope scales in the infrared with the \textit{Herschel Space Observatory} (\textit{Herschel})/HIFI guaranteed time CHESS key program \citep{Ceccarelli2010}, and para-D$_{2}$O tentatively \citep{Coutens2013b}. HDO has additionally been observed with SOFIA \citep{Parise2012}. Combined studies of data from the CHESS program, IRAM $30$-m and JCMT have detected water in the form of HDO, p-H$_{2}^{18}$O, o-H$_{2}^{18}$O and lone lines of p-H$_{2}^{17}$O, o-H$_{2}^{17}$O and HD$^{18}$O, but without the ability to spatially resolve the binary (smallest beam size probed was $10\farcs{4}$; \citealt{Coutens2012}).

Studies of water on the smallest  disc scales have so far been difficult. \citet{Persson2013} detected HDO and H$_{2}^{18}$O on scales of $\sim0\farcs{2}-5\farcs{0}$, thereby finally unambiguously disentangling the emission from sources A and B. Towards source A, the lines appear in emission. With the smallest beam size available, the emitting region seems to be marginally spatially resolved. This source size ($0\farcs{2}$ specifically) is then used to derive the column densities while accounting for the beam dilution with bigger beam sizes. There is little dependence on $T_{\text{ex}}$ in the $80$ to $300$~K range and the best estimate is $124\pm12$~K. The values of \citet{Persson2013} used in this paper are given in Table~\ref{tbl:abunvalues_B}. Towards source B, only one line is detected in absorption and column density estimates are not available. However, future work will obtain these column densities (A. Coutens priv. comm., which will also include a check on the values derived for source A).

\subsubsection{Carbon monoxide (CO), $m=28$}
\label{CO}
After the initial identification of IRAS~16293-2422 in the infrared, the first molecules that the region was studied in were $^{12}$CO, $^{13}$CO, C$^{18}$O, C$^{17}$O at radio wavelengths \citep{Young1986, Mundy1986, Menten1987, Wootten1987, Mundy1990}. Detections in the infrared have also been made with the Infrared Space Observatory (ISO)/Long Wavelength Spectrometer (LWS; \citealt{Ceccarelli1998a}) and with the \textit{Herschel}/HIFI guaranteed time CHESS key program \citep{Ceccarelli2010}. Studies were subsequently carried out at available increasing spatial resolutions \citep{Blake1994, Lis2002, Yen2008, Rao2009, Caux2011, Jorgensen2011}; and CO envelope abundances started to be modeled with jump profiles \citep{Schoier2002, Schoier2004}. $^{13}$C$^{18}$O and $^{13}$C$^{17}$O have also been detected on envelope scales \citep{Blake1994}. It has been shown that CO and its isotopologues are present on the large cloud and core scales in L1689N; and on the smaller circumbinary envelope, and disc-A and -B scales. It also traces the outflows from the system \citep{Walker1988, Mizuno1990}, as also solidified by the detection of CO$^{+}$ in ISO/LWS data (\citealt{Ceccarelli1998b}, although still remaining to be verified with ALMA observations). Most recently, studies with spatial resolution superseding that of PILS have been used to study the shocks and outflows stemming from source A as traced by carbon monoxide emission (\citealt{Kristensen2013}, based on the line labelled in \citealt{Baryshev2015}; \citealt{Girart2014, Favre2014b, vanderWiel2019}). Unfortunately, estimates of the column density are not provided in those works as too few lines are targeted at a time. \citet{Jacobsen2018a} developed sophisticated models of the binary IRAS~16293-2422 system based on $^{13}$CO, C$^{17}$O and C$^{18}$O $J=3-2$ observations. The latest derivation of the column density is presented in Appendix~\ref{columndens_IRAS16293B}.

\subsubsection{Molecular oxygen (O$_{2}$), $m=32$}
\label{O2}
\citet{Taquet2018} searched for molecular oxygen towards source B through the $^{16}$O$^{18}$O line at $234$~GHz, claiming at most a tentative detection.

\subsubsection{Hydroperoxyl (HO$_{2}$), $m=33$}
\label{HO2}
\citet{Taquet2018} searched for HO$_{2}$ towards source B and derived an upper limit.

\subsubsection{Hydrogen peroxide (H$_{2}$O$_{2}$), $m=34$}
\label{H2O2}
\citet{Taquet2018} searched for H$_{2}$O$_{2}$ towards source B, but were unable to derive an upper limit due to blending with other species.

\subsubsection{Formaldehyde (H$_{2}$CO), $m=30$}
\label{H2CO}
H$_{2}$CO, H$_{2}^{13}$CO, HDCO, H$_{2}$C$^{18}$O and D$_{2}$CO have been detected on envelope scales, and with both ortho- and para- versions showing self-absorption \citep{vanDishoeck1995, Ceccarelli1998c, Loinard2000, Chandler2005, Caux2011, vanderWiel2019} and towards both protostars, even as D$_{2}^{13}$CO \citep{Jorgensen2011}. A number of models have addressed the abundance of formaldehyde and its release via thermal desorption in the vicinity of the protostars \citep{Ceccarelli2000b, Ceccarelli2001, Schoier2002, Ceccarelli2003, Schoier2004}. H$_{2}$CO was detected on envelope scales in the infrared with the \textit{Herschel}/HIFI guaranteed time CHESS key program \citep{Ceccarelli2010}. Formaldehyde has also been shown to be associated with larger scale shocked regions \citep{Castets2001}, although it is significantly more abundant in the warm inner regions around the protostars \citep{Lindberg2017}. Individual lines of D$_{2}$CO and HDCO have been labelled in high spatial resolution observations of \citet{Baryshev2015} and \citet{Martin-Domenech2016}. Estimates of the column densities towards A and B have been provided in \citet{Rivilla2019}; however, corrections for optically thick emission have not been carried out. \citet{Persson2018} identified H$_{2}$CO, H$_{2}^{13}$CO, H$_{2}$C$^{17}$O, H$_{2}$C$^{18}$O, HDCO, HDC$^{18}$O, D$_{2}$CO, D$_{2}^{13}$CO towards source B and derived accurate column densities. Manigand et al. subm. identified H$_{2}$CO, H$_{2}^{13}$CO, H$_{2}$C$^{17}$O, H$_{2}$C$^{18}$O, HDCO, and D$_{2}$CO towards source A and derived accurate column densities.

\subsubsection{Methanol (CH$_{3}$OH), $m=32$}
\label{CH3OH}
CH$_{3}$OH, $^{13}$CH$_{3}$OH, CH$_{2}$DOH, CHD$_{2}$OH, CH$_{3}$OD and CD$_{3}$OH have been detected on envelope scales \citep{vanDishoeck1995, Parise2002, Parise2004, Kuan2004, Chandler2005, Caux2011}. On larger cloud and core scales, CH$_{3}$OH emission seems to arise from the material entrained by a wide-angle stellar wind stemming from the protostar and is estimated to be enhanced by a factor of $\sim500$ relative to the ambient gas \citep{Garay2002, Lis2016, Lindberg2017}. CH$_{3}$OH was detected on envelope scales in the infrared with the \textit{Herschel}/HIFI guaranteed time CHESS key program \citep{Ceccarelli2010}. CH$_{3}$OH and $^{13}$CH$_{3}$OH have been observed towards the two protostars individually \citep{Jorgensen2011}, even at spatial resolution superseding that of PILS -- lines of CH$_{3}$OH, $^{13}$CH$_{3}$OH, CH$_{2}$DOH have been labelled in \citet{Baryshev2015} and \citet{Martin-Domenech2016}. Towards source B, CH$_{3}$OH emission displays inverse P Cygni profiles, indicating infall \citep{Zapata2013}. Estimates of the column densities towards A and B have been provided in \citet{Rivilla2019}; however, corrections for optically thick emission have not been carried out. Additionally, CH$_{3}$OH has been argued to display abundance enhancements at the centrifugal barrier in both sources \citep{Oya2016, Oya2018}; however, these studies claim optically thin emission, which is likely not the case as analysed in \citet{Jorgensen2016, Jorgensen2018} based on CH$_{3}$OH, CH$_{2}$DOH and CH$_{3}$OD emission and accurate column densities towards source B. CH$_{3}^{18}$OH, CH$_{2}$DOH and CH$_{3}$OD have also been detected towards source A, like towards source B, and accurate column densities have been derived (Manigand et al. subm.).

\subsubsection{Ethanol (C$_{2}$H$_{5}$OH), $m=46$}
\label{C2H5OH}
g-C$_{2}$H$_{5}$OH was initially detected towards source A \citep{Huang2005} and then towards both sources \citep{Bisschop2008, Jorgensen2011}. \citet{Jorgensen2018} identified CH$_{3}$CH$_{2}$OH, a-a-CH$_{2}$DCH$_{2}$OH, a-s-CH$_{2}$DCH$_{2}$OH, a-CH$_{3}$CHDOH, a-CH$_{3}$CH$_{2}$OD, a-CH$_{3}^{13}$CH2OH, a-$^{13}$CH$_{3}$CH$_{2}$OH towards source B and derived accurate column densities. Manigand et al. subm. identified CH$_{3}$CH$_{2}$OH, a-a-CH$_{2}$DCH$_{2}$OH, a-s-CH$_{2}$DCH$_{2}$OH, a-CH$_{3}$CHDOH, a-CH$_{3}$CH$_{2}$OD towards source A and derived accurate column densities.

\subsubsection{Ketene/Ethenone (CH$_{2}$CO), $m=42$}
\label{CH2CO}
CH$_{2}$CO has been detected on envelope scales \citep{vanDishoeck1995, Caux2011} and towards the two sources individually \citep{Kuan2004, Bisschop2008, Jorgensen2011}. Inverse P Cygni profiles are seen near source B, unlike near source A, suggesting on-going infall and face-on orientation \citep{Pineda2012}. Lines of CH$_{2}$CO are also labelled in \citet{Majumdar2016}. \citet{Jorgensen2018} identified CH$_{2}$CO, $^{13}$CH$_{2}$CO, CH$_{2}^{13}$CO, CHDCO towards source B and derived accurate column densities. Manigand et al. subm. identified CH$_{2}$CO and CHDCO towards source A and derived accurate column densities.

\subsubsection{Formic acid (HCOOH), $m=46$}
\label{HCOOH}
HCOOH has been detected towards source B \citep{Cazaux2003, RemijanHollis2006} and towards A \citep{Jorgensen2011}, as has HCOOD \citep{Jorgensen2011}. \citet{Jorgensen2018} and Manigand et al. subm. identified t-HCOOH, t-H$^{13}$COOH, t-DCOOH, t-HCOOD towards sources B and A, and derived accurate column densities.

\subsubsection{Acetaldehyde (CH$_{3}$CHO), $m=44$}
\label{CH3CHO}
CH$_{3}$CHO was detected in its -A and -E forms towards both sources \citep{Cazaux2003, Bisschop2008, Jorgensen2011} and also on envelope scales \citep{Caux2011}. Its lines have also been labelled in \citet{Baryshev2015} and \citet{Martin-Domenech2016}, and mentioned in \citet{Girart2014}. \citet{Lykke2017, Jorgensen2018, Coudert2019} identified CH$_{3}$CHO, CH$_{3}$CDO, $^{13}$CH$_{3}$CHO, CH$_{3}^{13}$CHO and CH$_{2}$DCHO towards source B and derived accurate column densities; and CH$_{3}$CHO towards source A (Manigand et al. subm.).

\subsubsection{Methyl formate (HCOOCH$_{3}$), $m=60$}
\label{HCOOCH3}
HCOOCH$_{3}$ has been detected in its -A and -E forms towards both protostars individually \citep{Cazaux2003, Bottinelli2004, Kuan2004, Chandler2005, RemijanHollis2006, Shiao2010}. On circumbinary envelope scales, initially, only upper limits were available \citep{Remijan2003}, but then a detection was also made \citep{Caux2011}. DCOOCH$_{3}$ was tentatively detected with the TIMASSS survey \citep{Demyk2010}. Inverse P Cygni profiles are seen near source B, unlike near source A, suggesting on-going infall and face-on orientation \citep{Pineda2012, Favre2014b}. Its lines have also been labelled in \citet{Baryshev2015} and \citet{Martin-Domenech2016}. Additionally, HCOOCH$_{3}$ has been argued to display abundance enhancements at the centrifugal barrier in both sources \citep{Oya2016, Oya2018}; however, these studies claim optically thin emission, which is likely not the case as analysed in \citet{Jorgensen2018} based on CH$_{3}$OCHO, CH$_{3}$OCDO, CH$_{2}$DOCHO and CH$_{3}$O$^{13}$CHO emission towards source B. \citet{Manigand2019} and Manigand et al. subm. detected CHD$_{2}$OCHO, CH$_{2}$DOCHO, CH$_{3}$OCDO, CH$_{3}$O$^{13}$CHO and CH$_{3}$OCHO towards both protostars. Accurate column densities were derived in these works (\citealt{Jorgensen2018, Manigand2019}; Manigand et al. subm.).

\subsubsection{Acetic acid (CH$_{3}$COOH), $m=60$}
\label{CH3COOH}
CH$_{3}$COOH was detected in its -E form towards source B \citep{Cazaux2003}. Subsequently, in its -E and -A forms towards source A, but with only upper limits derived towards source B \citep{Shiao2010}. Higher spatial resolution observations secured its detection towards both sources \citep{Jorgensen2011}. On envelope scales, only upper limits have been derived \citep{Remijan2003}. CH$_{3}$COOH was again observed towards source B by \citet{Jorgensen2016} with an accurate column density being derived, and towards source A (Manigand et al. subm.).

\subsubsection{Glycolaldehyde (CH$_{2}$OHCHO), $m=60$}
\label{HCOCH2OH}
The first time detection towards both sources was obtained by \citet{Jorgensen2012}. The submitted paper of \citet{Zhou2018} erroneously claims the first time detection of vibrationally excited glycolaldehyde, as $v=1$ and $v=2$ lines have been detected towards both sources already in \citet{Jorgensen2012}. Column densities towards A and B have been derived in \citet{Rivilla2019} on $\sim1\arcsec$ scales. \citet{Jorgensen2016} identified CH$_{2}$OHCHO, CHDOHCHO, CH$_{2}$ODCHO, CH$_{2}$OHCDO, $^{13}$CH$_{2}$OHCHO, CH$_{2}$OH$^{13}$CHO towards source B and provided accurate column densities on $\sim0\farcs{5}$ scales; and CH$_{2}$OHCHO towards source A (Manigand et al. subm.).

\subsubsection{Ethylene glycol/Ethane-1,2-diol ((CH$_{2}$OH)$_{2}$), $m=62$}
\label{CH2OH_2}
A tentative detection towards source B has been reported in \citet{Jorgensen2012}. \citet{Jorgensen2016} firmly identified aGg\textquotesingle-ethylene glycol and gGg\textquotesingle-ethylene glycol towards source B and derived accurate column densities. The detection of ethylene glycol towards source A has been claimed by the submitted paper of \citet{Zhou2018}; however, this was already established earlier by the PILS team (Manigand et al. subm.). 

\subsubsection{Dimethyl ether/Methoxymethane (CH$_{3}$OCH$_{3}$), $m=46$}
\label{CH3OCH3}
CH$_{3}$OCH$_{3}$ was initially detected towards source B \citep{Cazaux2003}, but then towards both protostars \citep{Huang2005, Chandler2005, Jorgensen2011} and on envelope scales \citep{Caux2011}. CH$_{2}$DOCH$_{3}$ has also been detected on envelope scales \citep{Richard2013}. CH$_{3}$OCH$_{3}$ lines have been labelled in \citet{Baryshev2015} and in \citet{Martin-Domenech2016}. \citet{Jorgensen2018} and Manigand et al. subm. identified CH$_{3}$OCH$_{3}$, $^{13}$CH$_{3}$OCH$_{3}$, asym-CH$_{2}$DOCH$_{3}$, sym-CH$_{2}$DOCH$_{3}$ towards sources A and B, and derived accurate column densities.

\subsubsection{Acetone/Propan-2-one ((CH$_{3}$)$_{2}$CO), $m=58$}
\label{CH3_2_CO}
(CH$_{3}$)$_{2}$CO has been detected towards both sources by \citet{Jorgensen2011}, and once more by \citet{Lykke2017} and Manigand et al. subm. with accurate column densities being derived.

\subsubsection{Ethylene oxide (c-C$_{2}$H$_{4}$O), $m=44$}
\label{c_C2H4O}
Ethylen oxide was detected towards source B by \citet{Lykke2017} and an accurate column density was derived; towards source A (Manigand et al. subm.).

\subsubsection{Propanal/Propionaldehyde (C$_{2}$H$_{5}$CHO), $m=58$}
\label{C2H5CHO}
Propanal was detected towards sources B and A by \citet{Lykke2017} and Manigand et al. subm., and accurate column densities were derived.

\subsubsection{Vinyl alcohol (CH$_{2}$CHOH), $m=44$}
\label{CH2CHCO}
An upper limit for vinyl alcohol towards source B was derived by \citet{Lykke2017}.

\subsubsection{Methoxymethanol (CH$_{3}$OCH$_{2}$OH)}
\label{CH3OCH2OH}
Methoxymethanol has been detected towards source B and an accurate column density has been derived, while for source A an upper limit has been computed (Manigand et al. subm.).

\subsubsection{Trans ethyl methyl ether (t-C$_{2}$H$_{5}$OCH$_{3}$)}
\label{tC2H5OCH3}
Trans ethyl methyl ether has been detected towards source B and an accurate column density has been derived, while for source A an upper limit has been computed (Manigand et al. subm.).

\subsubsection{Glycine (NH$_{2}$CH$_{2}$COOH), $m=75$}
\label{NH2CH2COOH}
A search for glycine was carried towards IRAS~16293-2422 with IRAM $30$-m telescope yielding an upper limit based on the non-detection \citep{Ceccarelli2000c}. The latest derivation of an upper limit on the column density is presented in Appendix~\ref{columndens_IRAS16293B}.

\subsubsection{Butane-1,2-diol (HOCH$_{2}$(HO)CHCH$_{2}$CH$_{3}$), $m=90$}
\label{HOCH2_HO_CHCH2CH3)}
Six conformers of 1,2-butanediol were searched for towards source B based ALMA data, but only an upper limit of $1\times10^{13}$~cm$^{-2}$ has been derived \citep{Vigorito2018}.

\subsubsection{Hydroxide (OH$^{-}$), $m=17$}
\label{OH}
OH$^{-}$ is an anion, which has been detected on circumbinary envelope scales with ISO/LWS \citep{Ceccarelli1998a}, but remains to be conclusively confirmed. OD has also been discovered on these scales with SOFIA \citep{Parise2012}. OH has been observed at cm wavelengths and its emission appears to have variable flares \citep{Colom2016}.

\subsubsection{Formyl cation (HCO$^{+}$), $m=29$}
\label{HCOp}
HCO$^{+}$ cation has been detected on large cloud and core scales, and smaller circumbinary envelope scales \citep{Mizuno1990, vanDishoeck1995, Narayanan1998, Lis2002, Stark2004, Caux2011, Lis2016, Quenard2018b}. Its lines show blue-shifted self-absorption, which has been interpreted as a sign of expansion \citep{Choi1999}. DCO$^{+}$ has also been detected on these scales \citep{Wootten1987, vanDishoeck1995, Lis2002, Stark2004, Caux2011, Lindberg2017, Quenard2018b}, and so have H$^{13}$CO$^{+}$ \citep{vanDishoeck1995, Narayanan1998, Lis2002, Stark2004, Rao2009, Caux2011, Jorgensen2011, Quenard2018b} and HC$^{18}$O$^{+}$ \citep{Blake1994, vanDishoeck1995, Caux2011, Quenard2018b}. D$^{13}$CO$^{+}$ \citep{Caux2011, Quenard2018b} and HC$^{17}$O$^{+}$ have also been detected \citep{Caux2011}. HCO$^{+}$ and H$^{13}$CO$^{+}$ in the infrared was detected on envelope scales in the infrared with the \textit{Herschel}/HIFI guaranteed time CHESS key program \citep{Ceccarelli2010}. On the scale of individual envelopes, DCO$^{+}$ has been detected only towards A \citep{Jorgensen2011}. Most recently, the distribution of DCO$^{+}$ has been analysed by \citet{Murillo2018} on the scale of the individual protostars and the circumbinary envelope.

\subsubsection{Formyl radical (HCO), $m=29$}
\label{HCO}
The HCO radical has been detected in the TIMASSS survey and analysed by \citet{BacmannFaure2016}. Very recently it has been mapped on $1\arcsec$ scales towards both protostars, whereby blue-shifted absorption is seen towards both sources, implying that A is undergoing infall just like B \citep{Rivilla2019}.

\subsubsection{Hydrocarboxyl (HOCO$^{+}$), $m=45$}
\label{HOCO}
A detection of HOCO$^{+}$ has been claimed based on two lines by \citet{Majumdar2018} on envelope scales.

\subsection{N-bearing molecules}

\subsubsection{Ammonia (NH$_{3}$), $m=17$}
\label{NH3}
NH$_{3}$ is a symmetric molecule with very weak hyperfine (inversion) emission lines at radio frequencies. It was detected on large scales of the cloud and core in emission \citep{Menten1987, Wootten1987, Mundy1990, Mizuno1990} and in emission and absorption in its ortho- and para- forms with the \textit{Herschel}/HIFI guaranteed time CHESS key program \citep{Ceccarelli2010, Hily-Blant2010}. NH$_{2}$D was detected on smaller envelope scales \citep{vanDishoeck1995, Caux2011}. Para-NH$_{2}$D and ortho-NH$_{3}$ were also studied across L1689N based on \textit{Herschel}/HIFI data by \citet{Lis2016}. Their data clearly showed strong NH$_{3}$ emission towards the location of IRAS~16293-2422 suggesting NH$_{3}$ emission to be optically thick towards that position, and also subject to absorption from the foreground material. Para-NH$_{2}$D was not dedected towards IRAS~16293-2422 at those frequencies. Unfortunately, the field of view of their ALMA ACA mosaic, which included a ND$_{3}$ line, did not cover the position of IRAS~16293-2422.

Subsequently, at the moment, the only estimate of the quantity of ammonia in IRAS~16293-2422 available is that on the scales of its circumbinary envelope. \citet{Lis2016} do not provide an estimate of the column density of NH$_{3}$ towards the position of IRAS~16293-2422. \citet{Caux2011} do not derive the column density of NH$_{2}$D due to severe blending. Estimates from infrared NH$_{3}$ observations of \citep{Hily-Blant2010} span a range larger than an order of magnitude for a source size of $12-18\arcsec$. Observations of \citet{vanDishoeck1995} of NH$_{2}$D can also be used to calculate the column density of NH$_{3}$ within a range spanning an order of magnitude, as the appropriate D/H ratio to use is unclear (hence, a range of $0.05-0.005$ is considered in this work, e.g., \citealt{Persson2014}) for a source size of $20\arcsec$. \citet{Mundy1990} provided the only direct estimate of the column density of NH$_{3}$ for a source size of $20\arcsec$; however, this value may be too low due to optical thickness of ammonia emission. In fact, this value gives the lower limit of the range of ammonia column densities used in this work. Finally, the NH$_{3}$ column density estimate of \citet{Mizuno1990} is for a beam size of $40\arcsec$. If a source size of $20\arcsec$, representative of the circumbinary envelope, is assumed, then the column density of \citet{Mizuno1990} can be corrected for beam dilution. These four values are then used to obtain the range of ammonia column densities considered in this work (Table~\ref{tbl:abunvalues_B}).

\subsubsection{Hydrogen cyanide (HCN), $m=27$}
\label{HCN}
HCN lines show deep self-absorption \citep{Mizuno1990}, which has been observed to be blue-shifted and interpreted as a sign of expansion \citep{Choi1999}. Studies on large cloud and core scales have testified to the presence of H$^{13}$CN and DCN \citep{Lis2002}. Smaller envelope scales have been shown to contain HCN (including its self-absorption), H$^{13}$CN, HC$^{15}$N, DCN originating from the two protostars \citep{vanDishoeck1995, Kuan2004, Takakuwa2007, Caux2011, Jorgensen2011, Wampfler2014, vanderWiel2019}. Its envelope abundances can be reproduced with jump profiles \citep{Schoier2002}. The vibrationally excited state of HCN has only been observed towards A \citep{Huang2005}. Source A was checked for D$^{13}$CN emission, but not detected \citep{Wampfler2014}. HCN, H$^{13}$CN and DCN have also been observed on envelope scales in the infrared with the \textit{Herschel}/HIFI guaranteed time CHESS key program \citep{Ceccarelli2010, Rice2018}. Observations at spatial resolution superseding that of PILS show that H$^{13}$CN and HC$^{15}$N display inverse P Cygni profiles near source B, indicating infall, a small velocity gradient and an almost face-on orientation for the B-disc (\citealt{Zapata2013}, who analysed the lines labelled in \citealt{Baryshev2015}). HCN and DCN are covered in the data of \citet{Lis2016}, but, unfortunately, are not discussed towards the position of IRAS~16293-2422. The works of \citet{Zapata2013, Wampfler2014} targeted too few lines to estimate column densities.  The latest derivation of the column density is presented in Appendix~\ref{columndens_IRAS16293B}, which has also been mentioned in \citet{Rice2018}.

\subsubsection{Hydrogen isocyanide (HNC), $m=27$}
\label{HNC}
HNC, HN$^{13}$C and DNC have been detected on smaller envelope scales \citep{vanDishoeck1995, Caux2011} at abundances that are reproducible with jump profiles \citep{Schoier2002}. H$^{15}$NC was additionally detected towards source A \citep{Wampfler2014}. HNC was detected on envelope scales in the infrared with the \textit{Herschel}/HIFI guaranteed time CHESS key program \citep{Ceccarelli2010}. HNC and DNC are covered in the data of \citet{Lis2016}, but, unfortunately, are not discussed towards the position of IRAS~16293-2422. The works of \citet{Caux2011, Wampfler2014} targeted too few lines to estimate column densities. The latest estimate of the column density is presented in Appendix~\ref{columndens_IRAS16293B}.

\subsubsection{Nitroxyl/Azanone/Nitrosyl hydride (HNO), $m=31$}
\label{HNO}
A tentative detection of DNO was claimed in absorption towards source A \citep{Chandler2005}. \citet{Coutens2019b} derived an upper limit for HNO towards source B.

\subsubsection{Cyanic acid (HOCN), $m=43$}
\label{HOCN}
\citet{Marcelino2010} claimed a tentative detection of HOCN. This was not confirmed by \citet{Ligterink2017}.

\subsubsection{Fulminic acid/Oxidoazaniumylidynemethane (HCNO), $m=43$}
\label{HCNO}
\citet{Marcelino2010} reported a non-detection of HCNO.

\subsubsection{Isocyanic acid (HNCO), $m=43$}
\label{HNCO}
HNCO has been detected on the envelope scales \citep{vanDishoeck1995, Caux2011} and towards both protostars individually \citep{Bisschop2008, Jorgensen2011}. An upper limit for H$^{15}$NCO has been derived \citep{vanDishoeck1995}. HNC$^{18}$O was detected towards both, A and B \citep{Jorgensen2011}. A combined study of IRAM 30-m TIMASSS, APEX, \textit{Herschel}/HIFI CHESS and ALMA PILS data has been carried out by \citet{Hernandez-Gomez2019a} to infer column densities of HNCO for different components of the system. The authors argued that the emission in their observations is dominated by the nine times brighter source A; hence, the presented values are thought to be representative of the large-scale structures encompassing A (Section~$2.4$ of \citealt{Hernandez-Gomez2019a}). \citet{Coutens2016, Ligterink2017} detected DNCO and HN$^{13}$CO towards source B and derived accurate column densities for them, and derived upper limits for H$^{15}$NCO and HNC$^{18}$O. \citet{Martin-Domenech2017} detected HNC$^{18}$O towards source B and consequently, also derived a column density of HNCO. Manigand et al. subm. detected HNCO and DNCO towards source A and derived accurate column densities.

\subsubsection{Methyl cyanide/Acetonitrile (CH$_{3}$CN), $m=41$}
\label{CH3CN}
CH$_{3}$CN has been observed on envelope scales \citep{vanDishoeck1995, Caux2011, Andron2018} and then individually towards both protostars \citep{Cazaux2003, Bottinelli2004, Bisschop2008, Jorgensen2011}. CH$_{2}$DCN has been labelled in fig.~$1$ of \citet{Parise2004} of envelope-scale emission and detected towards both protostars by \citet{Jorgensen2011}. CH$_{3}^{13}$CN was detected towards source B by \citet{Bisschop2008}, and then towards both \citep{Jorgensen2011}. \citet{Calcutt2018a} detected methyl cyanide and five of its isotopologues ($^{13}$CH$_{3}$CN, CH$_{3}^{13}$CN, CH$_{3}$C$^{15}$N, CH$_{2}$DCN and CHD$_{2}$CN) towards both sources A and B, and derived accurate column densities.

\subsubsection{Vinyl cyanide/Acrylonitrile (C$_{2}$H$_{3}$CN), $m=53$}
\label{C2H3CN}
C$_{2}$H$_{3}$CN has been firmly detected, and CH$_{2}$CDCN tentatively, towards the two protostars individually \citep{Kuan2004, Huang2005}. \citet{Calcutt2018a} re-confirmed the detection of vinyl cyanide towards source B and derived an accurate column density, but only were able to derive an upper limit towards source A.

\subsubsection{Ethyl cyanide/Propionitrile (C$_{2}$H$_{5}$CN), $m=55$}
\label{C2H5CN}
C$_{2}$H$_{5}$CN was initially detected only towards source B \citep{Cazaux2003, Huang2005}, but then quickly towards both \citep{RemijanHollis2006, Jorgensen2011}. Subsequent observations with the Combined Array for Research in Millimeter-wave Astronomy (CARMA) yielded only upper limits \citep{Shiao2010}. Observations on envelope scales have also only yielded upper limits \citep{Remijan2003}. Ethyl cyanide was re-confirmed towards both sources by \citet{Calcutt2018a} with accurate column densities derived.

\subsubsection{Formamide (NH$_{2}$CHO), $m=45$}
\label{NH2CHO}
First detection was secured with TIMASSS in conjunction with newer IRAM $30$~m data on envelope scales \citep{Kahane2013}. \citet{Coutens2016} detected NH$_{2}$CHO in the $v=0$ and $v_{12}=1$ states, NH$_{2}$CDO, cis-NHDCHO, trans-NHDCHO, NH$_{2}^{13}$CHO towards source B and computed accurate column densities, as well as derived upper limits for $^{15}$NH$_{2}$CHO and NH$_{2}$CH$^{18}$O. Manigand et al. subm. detected NH$_{2}$CHO towards source A and derived an accurate column density.

\subsubsection{Methyl isocyanide (CH$_{3}$NC), $m=41$}
\label{CH3NC}
\citet{Calcutt2018b} detected CH$_{3}$NC towards source B and computed an accurate column density, and derived an upper limit towards source A.

\subsubsection{Methyl isocyanate (CH$_{3}$NCO), $m=57$}
\label{CH3NCO}
Methyl isocyanate was independently detected by \citet{Ligterink2017} and \citet{Martin-Domenech2017} towards source B, and towards source A by \citet{Ligterink2017}.

\subsubsection{Acetonitrile oxide (CH$_{3}$CNO), $m=57$}
\label{CH3CNO}
CH$_{3}$CNO was searched for by \citet{Ligterink2017}, but undetected.

\subsubsection{Methyl Cyanate (CH$_{3}$OCN), $m=57$}
\label{CH3OCN}
CH$_{3}$OCN was searched for by \citet{Ligterink2017}, but undetected.

\subsubsection{Cyanide (CN), $m=26$}
\label{CN}
CN has been detected on envelope scales \citep{vanDishoeck1995, Caux2011}.

\subsubsection{Cyanoacetylene (HC$_{3}$N), $m=51$}
\label{HC3N}
HC$_{3}$N was detected by \citet{vanDishoeck1995}, while \citet{Caux2011} additionally detected DC$_{3}$N. Its envelope abundances can be reproduced with jump profiles \citep{Schoier2002}. The emission from HC$_{3}$N seems to originate from both sources \citep{Kuan2004, Chandler2005}; however, \citet{Jorgensen2011} were able to secure a detection only towards source A. HC$_{3}$N and DC$_{3}$N were observed on envelope scales in TIMASSS \citep{JaberAl-Edhari2017}. \citet{Calcutt2018a} firmly detected HC$_{3}$N towards both protostars and derived accurate column densities.

\subsubsection{Cyanobutadiyne (HC$_{5}$N), $m=75$}
\label{HC5N}
HC$_{5}$N has been detected on envelope scales in TIMASSS \citep{JaberAl-Edhari2017}.

\subsubsection{Methylene amidogen (H$_{2}$CN), $m=28$}
\label{H2CN}
H$_{2}$CN has been detected towards both protostars \citep{Jorgensen2011}.

\subsubsection{Nitric oxide (NO), $m=30$}
\label{NO}
NO has been detected on envelope scales \citep{Caux2011}, and towards source B \citep{Ligterink2018b} with an accurate column density provided.

\subsubsection{Nitrous oxide (N$_{2}$O), $m=44$}
\label{N2O}
N$_{2}$O has been detected towards source B by \citet{Ligterink2018b} with an accurate column density provided.

\subsubsection{Hydroxylamine (NH$_{2}$OH), $m=33$}
\label{NH2OH}
An upper limit for NH$_{2}$OH towards source B has been derived by \citet{Ligterink2018b}.

\subsubsection{Methanimine (CH$_{2}$NH), $m=29$}
\label{CH2NH}
CH$_{2}$NH has been detected towards source B by \citet{Ligterink2018b} with an accurate column density provided.

\subsubsection{Methylamine (CH$_{3}$NH$_{2}$), $m=31$}
\label{CH3NH2}
An upper limit for CH$_{3}$NH$_{2}$ towards source B has been derived by \citet{Ligterink2018b}.

\subsubsection{Acetamide (CH$_{3}$C(O)NH$_{2}$), $m=59$}
\label{CH3CONH2}
Acetamide has been tentatively detected towards source B by \citet{Ligterink2018a}.

\subsubsection{N-methylformamide (CH$_{3}$NHCHO), $m=59$}
\label{CH3NHCHO}
An upper limit for N-methylformamide has been derived towards source B by \citet{Ligterink2018a}.

\subsubsection{Cyanamide (NH$_{2}$CN), $m=42$}
\label{NH2CN}
NH$_{2}$CN, NHDCN and NH$_{2}^{13}$CN have been detected towards source B by \citet{Coutens2018} with accurate column densities derived. H$_{2}^{15}$NCN has been tentatively detected, and an upper limit for H$_{2}$NC$^{15}$N has been derived towards source B \citep{Coutens2019a}.

\subsubsection{Imidogen (NH), $m=15$}
\label{NH}
The NH radical has been detected in absorption in the infrared with the \textit{Herschel}/HIFI guaranteed time CHESS key program \citep{Hily-Blant2010, Bacmann2010}. The ND radical has also been detected \citep{Bacmann2010}.

\subsubsection{Amidogen (NH$_{2}$), $m=16$}
\label{NH2m}
The NH$_{2}$ radical has been detected in absorption in the infrared with the \textit{Herschel}/HIFI guaranteed time CHESS key program \citep{Hily-Blant2010}.

\subsubsection{Diazenylium (N$_{2}$H$^{+}$), $m=29$}
\label{N2Hp}
N$_{2}$H$^{+}$ and N$_{2}$D$^{+}$ were detected at radio frequencies on large cloud and core scales, as well as envelope scales \citep{Castets2001, Lis2002, Lis2016}. N$_{2}$H$^{+}$ was also detected on envelope scales in the infrared with the \textit{Herschel}/HIFI guaranteed time CHESS key program \citep{Ceccarelli2010}. On the scale of individual envelopes, N$_{2}$D$^{+}$ is seen only towards A \citep{Jorgensen2011}. Most recently, the distribution of N$_{2}$D$^{+}$ has been analysed by \citet{Murillo2018} on the scale of the individual protostars and the circumbinary envelope.

\subsubsection{Glycolonitrile (HOCH$_{2}$CN), $m=57$}
\label{HOCH2CN}
Glycolonitrile was detected towards source B by \citet{Zeng2019}.

\subsubsection{Aminoacetonitrile (NH$_{2}$CH$_{2}$CN), $m=56$}
\label{NH2CH2CN}
An upper limit was derived towards source B by \citet{Zeng2019}.

\subsubsection{Formyl cyanide (CHOCN), $m=55$}
\label{CHOCN}
An upper limit was derived towards source B by \citet{Zeng2019}.

\subsubsection{Ethenimine (CH$_{2}$CNH), $m=41$}
\label{CH2CNH}
An upper limit was derived towards source B by \citet{Zeng2019}.

\subsubsection{Nitrous acid (HONO), $m=47$}
\label{HONO}
\citet{Coutens2019b} detected HONO towards source B and derived an accurate column density.

\subsubsection{Nitrogen dioxide (NO$_{2}$), $m=46$}
\label{NO2}
\citet{Coutens2019b} derived an upper limit for NO$_{2}$ towards source B.

\subsubsection{Nitrosyl cation (NO$^{+}$), $m=30$}
\label{NO+}
\citet{Coutens2019b} derived an upper limit for NO$^{+}$ towards source B.

\subsubsection{Nitric acid (HNO$_{3}$), $m=63$}
\label{HNO3}
\citet{Coutens2019b} derived an upper limit for HNO$_{3}$ towards source B.

\subsection{S-bearing molecules}

\subsubsection{Hydrogen sulphide (H$_{2}$S), $m=34$}
\label{H2S}
H$_{2}$S was detected on envelope scales \citep{Blake1994}, as was HDS \citep{vanDishoeck1995, Caux2011}. It was also observed towards the two protostars individually \citep{Chandler2005, Jorgensen2011}. On large cloud and core scales, H$_{2}$S is also observed, but at lower abundances than closer to the binary source \citep{Wakelam2004a}. H$_{2}$S was detected on envelope scales in the infrared with the \textit{Herschel}/HIFI guaranteed time CHESS key program \citep{Ceccarelli2010}. \citet{Baryshev2015} reported a line of H$_{2}$S in absorption from the foreground cloud. HDS and HD$^{34}$S were detected towards source B by \citet{Drozdovskaya2018} with accurate column densities provided.

\subsubsection{Carbon monosulphide (CS), $m=44$}
\label{CS}
C$^{32}$S was the second molecule studied in IRAS~16293-2422. Its emission line profile showed strong self-absorption and prominent asymmetry, which was interpreted as a signature of infall \citep{Walker1986,vanDishoeck1995}. C$^{34}$S was observed soon thereafter \citep{Menten1987}. CS, like CO, is present on the large cloud and core scales; and the smaller circumbinary envelope and disc-A and -B scales \citep{Mizuno1990, Walker1993, Narayanan1998, Blake1994, TakakuwaKamazaki2011, Caux2011, Jorgensen2011, vanderWiel2019}. It also traces the outflows \citep{Walker1990, Girart2014}, and shocks, as solidified by the detection of vibrationally-excited lines \citep{Walker1994, Blake1994}. Its envelope abundance has also been modeled with jump profiles \citep{Schoier2002}. $^{13}$CS, C$^{33}$S and C$^{34}$S were detected on smaller envelope scales \citep{Caux2011, Jorgensen2011, Favre2014b}. CS and C$^{34}$S were detected on envelope scales in the infrared with the \textit{Herschel}/HIFI guaranteed time CHESS key program \citep{Ceccarelli2010}. C$^{34}$S, C$^{33}$S and C$^{36}$S were detected towards source B by \citet{Drozdovskaya2018} with accurate column densities derived.

\subsubsection{Acetylene sulfide (C$_{2}$S), $m=56$}
C$_{2}$S has been detected on envelope scales \citep{Caux2011}.

\subsubsection{Carbonyl sulphide (OCS), $m=60$}
\label{OCS}
OCS, OC$^{34}$S, O$^{13}$CS and $^{18}$OCS have been detected on envelope scales \citep{Blake1994, vanDishoeck1995, Huang2005, Caux2011}. OCS has subsequently been detected towards both protostars individually, but O$^{13}$CS only towards A \citep{Jorgensen2011}. Additionally, OCS has been argued to display abundance enhancements at the centrifugal barrier in both sources \citep{Oya2016, Oya2018}; however, these studies claim optically thin emission, which is likely not the case as analysed in \citet{Drozdovskaya2018}. OCS in its ground and vibrationally excited states, as well as O$^{13}$CS, OC$^{34}$S, OC$^{33}$S, and $^{18}$OCS were detected towards source B by \citet{Drozdovskaya2018} with accurate column densities derived.

\subsubsection{Sulphur monoxide (SO), $m=48$}
\label{SO}
Initially, SO was only detected towards source A and proposed to be associated with outflows; while its non-detection towards source B indicated an abundance at least a factor of $10$ lower in comparison \citep{Mundy1992}. The detection towards A was verified several times \citep{Huang2005, Chandler2005}. SO and its vibrationally excited state, $^{34}$SO and S$^{18}$O were observed on circumbinary envelope scales \citep{Blake1994, Wakelam2004a, Caux2011}. SO was detected on envelope scales in the infrared with the \textit{Herschel}/HIFI guaranteed time CHESS key program \citep{Ceccarelli2010}. SO was finally detected towards both protostars, but $^{33}$SO and $^{34}$SO only towards A \citep{Jorgensen2011}. SO is labelled in \citet{Baryshev2015} for source B. \citet{Lindberg2017} also demonstrated its presence on the large cloud scales. The presence of SO was re-confirmed for source B by \citet{Drozdovskaya2018} with an accurate column density derived.

\subsubsection{Sulphur dioxide (SO$_{2}$), $m=64$}
\label{SO2}
SO$_{2}$ has been detected on envelope scales \citep{Blake1994, vanDishoeck1995, Caux2011} and large cloud and core scales \citep{Wakelam2004a} at lower abundances. Its vibrationally excited state was detected towards both inner regions of sources A and B, but $^{34}$SO$_{2}$ was only observed towards A \citep{Huang2005, Chandler2005}. SO$_{2}$ was detected on envelope scales in the infrared with the \textit{Herschel}/HIFI guaranteed time CHESS key program \citep{Ceccarelli2010}. Subsequently, SO$_{2}$ was observed towards both protostars, but $^{33}$SO$_{2}$, $^{34}$SO$_{2}$, SO$^{18}$O and SO$^{17}$O only towards A \citep{Jorgensen2011}. $^{33}$SO$_{2}$, $^{34}$SO$_{2}$ are labelled in \citet{Baryshev2015} for source B. The presence of SO$_{2}$ and $^{34}$SO$_{2}$ were re-confirmed for source B by \citet{Drozdovskaya2018} with accurate column densities derived.

\subsubsection{Thioformaldehyde (H$_{2}$CS), $m=46$}
\label{H2CS}
H$_{2}$CS, H$_{2}^{13}$CS, H$_{2}$C$^{34}$S and HDCS have been detected on envelope scales \citep{Blake1994, Huang2005, Caux2011}. Towards the two protostars individually, H$_{2}$CS and HDCS have been observed, but H$_{2}$C$^{34}$S and D$_{2}$CS only towards A \citep{Jorgensen2011}. H$_{2}$CS has been argued to trace the envelope and the A- and B-discs \citep{Oya2016, Oya2018}, which collaborates nearly optically thin emission claimed by \citet{Drozdovskaya2018}. The presence of H$_{2}$CS and HDCS were re-confirmed for source B by \citet{Drozdovskaya2018} with accurate column densities derived, and upper limits derived for H$_{2}$C$^{34}$S and D$_{2}$CS.

\subsubsection{Methyl mercaptan/Methanethiol (CH$_{3}$SH), $m=48$}
\label{CH3SH}
CH$_{3}$SH has been detected on circumbinary envelope scales \citep{Majumdar2016} and towards source B \citep{Drozdovskaya2018} with an accurate column density derived. An upper limits for CH$_{3}$SD has been derived towards source B \citep{Zakharenko2019a}.

\subsubsection{Ethyl mercaptan/Ethanethiol (C$_{2}$H$_{5}$SH), $m=62$}
\label{C2H5SH}
Upper limits for gauche- and anti-C$_{2}$H$_{5}$SH, and gauche-C$_{2}$H$_{5}^{34}$SH, were derived towards source B by \citet{Drozdovskaya2018}.

\subsubsection{Disulphur (S$_{2}$), $m=64$}
\label{S2}
An upper limit for S$_{2}$ on the envelope scale has been reported based on APEX data by \citet{Martin-Domenech2016}. An upper limit has also been derived towards source B by \citet{Drozdovskaya2018}.

\subsubsection{Trisulphur (S$_{3}$), $m=96$}
\label{S3}
An upper limit has been derived towards source B by \citet{Drozdovskaya2018}.

\subsubsection{Tetrathietane (S$_{4}$), $m=128$}
\label{S4}
An upper limit has been derived towards source B by \citet{Drozdovskaya2018}.

\subsubsection{Disulphanide (HS$_{2}$), $m=65$}
\label{HS2}
An upper limit for HS$_{2}$ on the envelope scale has been reported based on APEX data by \citet{Martin-Domenech2016}. An upper limit has also been derived towards source B by \citet{Drozdovskaya2018}.

\subsubsection{Disulphane (H$_{2}$S$_{2}$), $m=66$}
\label{H2S2}
An upper limit for H$_{2}$S$_{2}$ on the envelope scale has been reported based on APEX data by \citet{Martin-Domenech2016}. An upper limit has also been derived towards source B by \citet{Drozdovskaya2018}.

\subsubsection{Sulphur mononitride (NS), $m=46$}
\label{NS}
NS is detected towards source A \citep{Chandler2005} and on envelope scales \citep{Caux2011}. An upper limit was derived towards source B by \citet{Drozdovskaya2018}.

\subsubsection{Thioformyl (HCS), $m=45$}
\label{HCS}
HCS has been detected on envelope scales \citep{Blake1994, vanDishoeck1995, Caux2011}, also solely towards source A \citep{Jorgensen2011}. An upper limit has been derived towards source B by \citet{Drozdovskaya2018}.

\subsubsection{Thioformylium (HCS$^{+}$), $m=45$}
\label{HCSp}
DCS$^{+}$ has been tentatively detected towards source A \citep{Jorgensen2011}. Upper limits for HCS$^{+}$, DCS$^{+}$ and HC$^{34}$S$^{+}$ have been derived towards source B by \citet{Drozdovskaya2018}.

\subsubsection{Disulphur monoxide (S$_{2}$O), $m=80$}
\label{S2O}
An upper limit for S$_{2}$O was derived towards source B by \citet{Drozdovskaya2018}.

\subsubsection{Disulphur dioxide (S$_{2}$O$_{2}$), $m=96$}
\label{S2O2}
An upper limit for cis-S$_{2}$O$_{2}$ was derived towards source B by \citet{Drozdovskaya2018}.

\subsubsection{Thiohydroxymethylidyne (HSC), $m=45$}
\label{HSC}
An upper limit for HSC was derived towards source B by \citet{Drozdovskaya2018}.

\subsubsection{Thioethenylidene (CCS), $m=56$}
\label{CCS}
An upper limit for CCS was derived towards source B by \citet{Drozdovskaya2018}.

\subsubsection{Thioketene (H$_{2}$C$_{2}$S), $m=58$}
\label{H2C2S}
An upper limit for H$_{2}$C$_{2}$S was derived towards source B by \citet{Drozdovskaya2018}.

\subsubsection{c-C$_{2}$H$_{4}$S, $m=60$}
\label{cC2H4S}
An upper limit for c-C$_{2}$H$_{4}$S was derived towards source B by \citet{Drozdovskaya2018}.

\subsubsection{Thiocyanate (NCS), $m=58$}
\label{NCS}
An upper limit for NCS was derived towards source B by \citet{Drozdovskaya2018}.

\subsubsection{Isothiocyanic acid (HNCS), $m=59$}
\label{HNCS}
An upper limit for a-type HNCS was derived towards source B by \citet{Drozdovskaya2018}.

\subsubsection{Thiocyanic acid (HSCN), $m=59$}
\label{HSCN}
An upper limit for HSCN was derived towards source B by \citet{Drozdovskaya2018}.

\subsection{Other species}

\subsubsection{Hydrogen chloride (HCl), $m=36$}
\label{HCl}
HCl and H$^{37}$Cl were detected on envelope scales in the infrared with the \textit{Herschel}/HIFI guaranteed time CHESS key program \citep{Ceccarelli2010} and with the CSO \citep{Peng2010}.

\subsubsection{Methyl chloride (CH$_{3}$Cl), $m=50$}
\label{CH3Cl}
CH$_{3}$Cl and CH$_{3}^{37}$Cl were detected towards source B by \citet{Fayolle2017} with accurate column densities derived.

\subsubsection{Fluoromethane (CH$_{3}$F), $m=34$}
\label{CH3F}
An upper limit for CH$_{3}$F was derived towards source B by \citet{Fayolle2017}.

\subsubsection{Silicon monoxide (SiO), $m=44$}
\label{SiO}
SiO and $^{29}$SiO were detected on envelope scales \citep{Blake1994, Ceccarelli2000a, Caux2011} at abundances reproducible with jump profiles \citep{Schoier2002}. On large cloud and core scales, SiO emission seems to arise from the interaction of the outflowing gas and the dense ambient clump material, the outflow lobe itself and the central region of the lobe \citep{Hirano2001, Wakelam2004a}. It has been claimed to be consistent with C-type shock profiles and estimated to be enhanced by a factor of $\sim330$ relative to the ambient gas \citep{Garay2002, Girart2014}. SiO has been suggested to be masing at an off-source position between the two protostars \citep{RemijanHollis2006, Rao2009}. SiO has been observed towards the two protostars individually by \citet{Jorgensen2011} and \citet{vanderWiel2019}. \citet{Oya2018} address SiO emission near source B as a tracer of the outflow launch, but could not rule out the influence of A on B.

\subsubsection{Silicon monosulphide (SiS), $m=60$}
\label{SiS}
SiS and $^{29}$SiS were detected on envelope scales \citep{Blake1994}. Si$^{34}$S was tentatively detected towards source A \citep{Chandler2005}.

\subsubsection{Phosphorus mononitride (PN), $m=45$}
\label{PN}
A line of PN has been labelled on envelope scales \citep{Caux2011}. The latest derivation of an upper limit on the column density is presented in Appendix~\ref{columndens_IRAS16293B}.

\subsubsection{Trihydrogen (H$_{3}^{+}$), $m=3$}
\label{H3}
On cloud scales, H$_{2}$D$^{+}$ has been detected \citep{Stark2004}. Ortho- and para-H$_{2}$D$^{+}$ were detected with APEX in emission and with SOFIA in absorption, respectively \citep{Brunken2014}. Ortho-D$_{2}$H$^{+}$ was now also found with SOFIA in absorption \citep{Harju2017}.

\subsection{Detected atoms}
\label{atoms}
Several atomic lines have also been detected on circumbinary envelope scales towards this source, which include OI with the Kuiper Airborn Observatory \citep{Ceccarelli1997}, CI with the JCMT \citep{Ceccarelli1998a} and CII with ISO/LWS \citep{Ceccarelli1998a}.

\subsection{Hydrocarbons}

\subsubsection{Methylidyne (CH), $m=13$}
\label{CH}
CH has been observed in absorption in the infrared with the \textit{Herschel}/HIFI guaranteed time CHESS key program on the large cloud and circumbinary envelope scales \citep{Bottinelli2014}.

\subsubsection{Ethynyl (C$_{2}$H), $m=25$}
\label{C2H}
C$_{2}$H and C$_{2}$D have been detected on envelope scales \citep{vanDishoeck1995, Caux2011}. C$_{2}$H was detected and its distribution studied by \citet{Murillo2018} and \citet{vanderWiel2019} on the scale of the individual protostars and the circumbinary envelope.

\subsubsection{Cyclopropanediylidenyl (c-C$_{3}$H), $m=37$}
\label{cC3H}
c-C$_{3}$H has been detected on envelope scales \citep{Caux2011}.

\subsubsection{Cyclopropenylidene (c-C$_{3}$H$_{2}$), $m=38$}
\label{cC3H2}
c-C$_{3}$H$_{2}$ has been detected on cloud scales \citep{Lindberg2017}, envelope scales \citep{vanDishoeck1995, Caux2011} and on the scales of the individual envelopes \citep{Kuan2004, Majumdar2017}. An upper limit on C$_{3}$HD was reported by \citet{vanDishoeck1995} and a detection was secured by \citet{Caux2011} on the scales of the circumbinary envelope. On the scales of individual envelopes, c-C$_{3}$H$_{2}$ and c-C$_{3}$HD have now also been observed \citep{Majumdar2017}. c-C$_{3}$D$_{2}$ has been tentatively labelled in \citet{Martin-Domenech2016}. Most recently, its distribution has been analysed by \citet{Murillo2018} on the scale of the individual protostars and the circumbinary envelope.

\subsubsection{Butadiynyl (C$_{4}$H), $m=49$}
\label{C4H}
C$_{4}$H has been detected on envelope scales \citep{Caux2011}.

\subsubsection{Propyne (CH$_{3}$CCH), $m=40$}
\label{CH3CCH}
CH$_{3}$CCH was detected on envelope scales \citep{vanDishoeck1995, Caux2011, Andron2018} and towards source B \citep{Cazaux2003}. CH$_{2}$DCCH has also been observed on envelope scales \citep{Caux2011}. Calcutt et al. subm. have detected and derived accurate column densities of CH$_{3}$CCH towards both protostars, as well as upper limits for CH$_{3}^{13}$CCH, $^{13}$CH$_{3}$CCH, CH$_{3}$C$^{13}$CH, CH$_{3}$CCD, CH$_{2}$DCCH towards both protostars.

%%%%%%%%%%%%%%%%%%%%%%%%%%%%%%%%%%%%%%%%%%%%%%%%%%%%%%%%%%%%%%%%%%%%%%%%%%%%%%%
\bsp % ``This paper has been produced using the ...''

\label{lastpage}

\end{document}